\documentclass[]{aastex631}
\usepackage{amsmath}
\usepackage{booktabs}
\usepackage{array}
\newcolumntype{P}[1]{>{\raggedright}p{#1}} 

\begin{document}

\title{Asteroidal activity amongst meteor datasets: Confirmed new ``rock-comet'' stream and search for a tidal disruption signature}

\correspondingauthor{Patrick M. Shober}
\email{patrick.m.shober@nasa.gov}

\author[0000-0003-4766-2098]{Patrick M. Shober}
\affiliation{NASA Astromaterials Research and Exploration Science Division, Johnson Space Center, Houston TX 77058, USA}



\begin{abstract} 
Asteroid activity (e.g., thermo-mechanical breakdown, impacts, rotational shedding, tidal disruption, etc.) can inject meteoroids into near-Earth space and leave detectable signatures in orbit catalogs. We searched for such recent signatures using orbit-similarity statistics and explicit null-hypothesis testing applied to shower-removed, asteroidal video-meteor datasets. Our sample comprises 235{,}271 meteors and fireballs from four all-sky video networks (GMN, CAMS, EDMOND, and SonotaCo). For meteors we use the geocentric dissimilarity criterion $D_N$ and construct KDE-based sporadic null realizations to evaluate (i) global cumulative similarity distributions and (ii) localized $D_N$-conditioned ($D_N<0.015$) pair-excess maps in the $(U,\lambda_\odot)$ plane; we additionally apply DBSCAN ($\epsilon=0.03$, $\mathrm{min\_samples}=2$) to isolate the coherent, statistically significant structures. We find no survey-consistent, stream-like signature in the Earth-like, low-inclination region expected for a distinct \emph{recent} tidal-disruption family; instead, significant-bin membership implies, under our adopted detection thresholds and binning, a conservative combined upper limit of $\leq 53/235{,}271$ ($\leq 2.3\times10^{-4}$) for sporadic asteroidal meteors plausibly attributable to a detectable recent tidal-disruption-like contribution. In contrast, we confirm the detection of a new diffuse southern Virginid-region stream: GMN exhibits a local z-score of 6.32 relative to the KDE-null mean in the $U-\lambda_\odot$ phase space (global significance of 5.3~$\sigma$), with weaker supporting excess in SonotaCo and EDMOND. DBSCAN isolates $N=282$ members (243 GMN plus additional SonotaCo, CAMS, and EDMOND) on a low-perihelion, asteroidal orbit ($q=0.22\pm0.01$ au, $i=12.3^{\circ}\pm1.8^{\circ}$, $T_J=4.6\pm0.3$) consistent with near-Sun thermo-mechanical ``rock-comet'' activity. 
\end{abstract}

\keywords{Meteoroids(1040) --- Meteors(1041) --- Near-Earth objects(1092) --- Close encounters(255)}

\section{Introduction} \label{sec:intro}

The physical evolution of small bodies in the inner solar system is driven by several processes (solar heating, impacts, radiative spin-state torques, space weathering, close planetary encounters, etc.) that can modify bodies both physically and chemically \citep{chapman2004space,michel2015asteroids}. This physical evolution sometimes results in observable activity amongst asteroids (dust, meteoroid, or gas ejection), often grouped under the umbrella of ``active asteroids.'' These bodies, whose transient comae, tails, or fragment trains demonstrate that cometary ice sublimation is not the only pathway to producing meteoroids and maintaining the near-Earth dust environment \citep[e.g.,][]{hsieh2006population,jewitt2012active}. This activity can leave two complementary classes of signatures: (i) short-lived, dynamically young debris structures that retain orbital coherence (meteoroid streams and other excess-similarity clusters), and (ii) long-term, population-level distortions in the orbital and compositional distributions.

Meteor observations act as a uniquely sensitive probe of weak or episodic mass-loss because they directly sample the debris delivered to Earth, including cases where the parent has faded, evolved onto a less active state, or disrupted entirely. While most major meteor showers are linked to comets and sustained volatile-driven dust production, key exceptions establish that rocky bodies can supply substantial meteoroid streams. The Geminids, dynamically associated with near-Earth asteroid (3200)~Phaethon, remain the archetype of a high-mass, asteroid-associated stream. Phaethon exhibits recurrent near-perihelion brightening and tail development, which have motivated ``rock-comet'' interpretations in the near-Sun environment \citep[e.g.,][]{jewitt2010activity,jewitt2013dust}. The physical driver of Phaethon-like activity remains debated, with proposed roles for thermal dehydration and mineral decomposition with associated gas release \citep{maclennan2024thermal}; however, present-day perihelion brightening is consistent with sodium photoemission
near 589~nm (i.e., no dust) \citep{zhang2023sodium}. Regardless of the dominant mechanism in any individual case, these observations emphasize that non-ice, thermally driven mass-loss pathways (e.g., thermal decomposition/dehydration with gas release and thermomechanical breakdown) can inject fresh debris into near-Earth space and potentially generate large, dynamically young streams.

Several physical mechanisms can plausibly generate asteroidal meteoroids at rates sufficient to seed detectable short-term structure. (i)~Thermal fatigue driven by diurnal and seasonal temperature cycling can fracture rocks and rapidly generate regolith on airless bodies \citep{molaro2012rates,delbo2014thermal}. In situ observations at Bennu show boulder morphologies consistent with fatigue-driven exfoliation, supporting thermally induced breakdown as an active surface process on near-Earth asteroids \citep{molaro2017thermally,molaro2020situ,molaro_2020JGRE}. There is also direct evidence for thermally-induced fragmentation in meteoroids based on observations of ``meteor clusters'' \citep{capek2010thermal,capek2022ejection,ashimbekova2025towards}. At very small perihelia, dehydration and mineral decomposition can further enhance mass loss by releasing gas species and elevating pore pressures \citep{maclennan2024thermal}. (ii)~Impacts can directly eject particles and refresh surfaces; at Bennu, episodic particle ejection has been observed and meteoroid impacts are a plausible driver for at least a subset of events \citep{lauretta2019episodes,bottke2020meteoroid}. (iii)~Rotational evolution driven by radiative torques can spin up small, weak bodies toward surface shedding, fission, or disruption \citep{rubincam2000radiative,walsh2008rotational}, providing a pathway to meteoroid production even in the absence of volatiles. (iv)~Close encounters with terrestrial planets can tidally distort rubble piles, induce mass shedding, or catastrophically disrupt gravitational aggregates, producing fragment trains with initially similar orbits \citep{richardson1998tidal,binzel2010earth,zhang2020tidal}. In practice, multiple mechanisms may operate on the same object across its dynamical lifetime, and several produce low ejection speeds that are observationally difficult to distinguish in meteor datasets without using orbital-phase-space localization and stringent statistical null tests.

Near-Sun processing and disruption are also implicated by population-level signatures that are not, by themselves, discrete ``streams.'' Debiased NEO models show a depletion of low-albedo objects at small perihelion distances, interpreted as ``super-catastrophic'' disruption of a substantial fraction of asteroids once they reach sufficiently small \(q\) \citep{granvik2016super}. Related work connecting SOHO, fireball, and meteor constraints argues that efficient near-Sun disruption can supply millimeter-scale dust and influence the meteoroid environment inside \(\sim\)0.2~au \citep{wiegert2020supercatastrophic}. Independent lines of evidence from meteorite-fall statistics also indicate strong perihelion-dependent survivability for carbonaceous materials, consistent with thermal fragmentation operating as an effective filter on larger meteoroids at small \(q\) \citep{shober2025perihelion}. Together, these results motivate searches for ``orphan'' or weakly parented structures: debris produced recently enough to retain some orbital coherence, but potentially lacking an obvious parent due to fading, fragmentation, or complete disintegration.

Tidal disruption during encounters with Earth and Venus provides a particularly specific, testable hypothesis for the recent production of debris. Numerical simulations demonstrate that close planetary flybys can strongly reshape or disrupt weak gravitational aggregates \citep[e.g.,][]{richardson1998tidal,zhang2020tidal}. However, identifying tidal-disruption ``families'' among NEAs is expected to be difficult because orbital coherence is short-lived: synthetic families produced by catastrophic Earth encounters remain statistically detectable for only a few kyr to \(\sim10^4\)~yr, with coherence timescales of order \(10^4\)~yr \citep{schunova2014properties,shober2025decoherence}. Encounter physics is nonetheless supported by independent evidence that planetary flybys can refresh asteroid surfaces \citep{binzel2010earth}. Most recently, \citet{granvik2024tidal} reported an overabundance of NEAs with perihelia clustered near the orbital distances of Earth and Venus that cannot be explained by survey bias or standard orbital evolution, and interpreted this pattern as a signature of tidal disruption during close encounters. If such events are frequent, they should contribute fragments on Earth- and Venus-like perihelia with low inclinations and low encounter speeds, at least initially, offering a potential signature in both NEA and meteoroid orbital distributions. This question is especially timely ahead of the 2029 April 13 close approach of (99942)~Apophis, which is motivating coordinated ground-based characterization campaigns and dedicated spacecraft investigations (e.g., OSIRIS-APEX and ESA's RAMSES concept) that can directly test predictions for encounter-driven tidal perturbations \citep[e.g.,][]{reddy2022apophis,dellagiustina2023osiris,morelli2024initial}.

Whether an analogous encounter-driven tidal signature should appear in meteoroid and fireball orbits is not obvious. Even if tidal disruption produces large numbers of fragments with Earth- and Venus-like perihelia, the orbital coherence of such debris is expected to be short-lived and likely fragile: NEO-family simulations predict coherence times of order \(10^4-10^5\)~yr and detectability windows of only a few kyr for catastrophic events \citep{pauls2005decoherence,schunova2014properties,shober2025decoherence}. Population-level studies using meteorite-fall and fireball-scale datasets have found that apparent orbital associations are consistent with random chance alignments once sample size and orbit uncertainties are accounted for \citep{pauls2005decoherence,shober2025decoherence,shober2025determining}, implying that any contribution from very recent tidal disruption at centimeter--meter scales is, at most, a minute fraction of the impactor population. Detecting (or tightly constraining) such a weak recent signature is therefore more naturally suited to large meteor orbit databases (i.e., dust-sized meteoroids), which provide orders of magnitude more orbits than current fireball samples.

This paper addresses two linked questions. First, do modern meteor orbit catalogs contain statistically significant, localized excess orbit similarity that is indicative of recent asteroidal activity after strict removal of established showers and cometary interlopers?  Second, is any detected excess consistent with an encounter-driven tidal-disruption fingerprint associated with Earth and Venus? We address these questions using a large multi-network video meteor database together with a parallel analysis of NEA orbits.

\section{Data} \label{sec:data}
This work uses two meteor/fireball datasets for two distinct purposes. The first is a large four–video-network sample (N=235{,}271 after shower/cometary component removal) used for the orbit-similarity statistics and localized (U,$\lambda_\odot$) pair-excess mapping. This large meteor sample is taken from the European viDeo Meteor Observation Network Database (EDMOND), the Cameras for Allsky Meteor Surveillance (CAMS), the Global Meteor Network (GMN), and the SonotaCo network. Although EDMOND is not a single observing network (it aggregates data from multiple European programs), the other three conduct direct sky observations. This subset was used to search for very ``recent'' asteroidal activity (within last $\sim$50~kyr) and is comprised of 122{,}943 GMN, 60{,}733 CAMS, 30{,}341 EDMOND, and 21{,}254 SonotaCo observations. 

The second dataset used in this study, used for searching for ``long-term'' activity signatures (last several millions of years), is a separate multi-source compilation restricted to the observation-debiased regime, i.e., only large/bright meteors for which the limiting-magnitude observation bias is removed. In total, this smaller dataset comprises 18{,}012 events collected by GMN, CAMS, EDMOND, FRIPON \citep{colas2020fripon}, European Fireball Network \citep{borovivcka2022_one}, Global Fireball Observatory \citep{devillepoix2019observation}, US Government sensor (USG) detections\footnote{\url{https://cneos.jpl.nasa.gov/fireballs/}}, and recovered meteorite falls with orbits \citep{jenniskens2025review}. This data is used only for perihelion-distribution comparisons (Figs.~\ref{fig:g16_comparison} and \ref{fig:q_distro_normalized}), which require removing observational biases. 


\subsection{EDMOND}
EDMOND is a collaborative international database and a significant resource for recording and analyzing European meteor activity. EDMOND integrates data from national organizations such as BOAM (Base des Observateurs Amateurs de Météores) in France, CEMeNt (Central European Meteor Network), HMN (Hungarian Meteor Network), IMTN (Italian Meteor and TLE Networks), PFN (Polish Fireball Network), SVMN (Slovak Video Meteor Network), and UKMON (UK Meteor Observation Network). Additional observers from Bosnia, Serbia, Ukraine, and other countries also contribute to the database, which has grown significantly since its inception \citep{kornos2014edmond}. The two primary software systems used include MetRec (created by Sirko Molau; \citealt{molau1999meteor}) and the UFO Tools software developed by SonotoCo to detect single-station meteors \citep{sonotaco2009meteor}. The single-station observations output by each software are then combined via the UFOOrbit software for orbital solutions and compiled by EDMOND. Currently, EDMOND contains over 7 million individual meteor detections, producing more than 480,000 computed orbits spanning from 2000 to 2023. All data are publicly accessible online\footnote{\url{https://www.meteornews.net/edmond/edmond/edmond-database/}}, fostering an open repository of European meteor observations.

\subsection{CAMS}
Launched in 2010, the CAMS network was established to validate known minor meteor showers and discover new ones by reconstructing meteoroid orbits with multi-station video setups \citep{jenniskens2011cams}. CAMS currently operates over 560 low-light video cameras dispersed across nine countries, each surveying the sky above 31\,$^{\circ}$ in elevation, with the capacity to detect meteors down to around magnitude +4. Each station employs an array of Watec 902 H2 Ultimate cameras (generally around 20 per site), featuring a stellar limiting magnitude of about +5.4. Automatically processed streaks are matched across stations via a coincidence algorithm, enabling precise triangulation of atmospheric trajectories and velocities using the Borovička straight-line least squares method \citep{Borovicka_1990BAICz,jenniskens2016established}. This software accounts for deceleration owing to aerodynamic drag and gravitational influences, deriving robust pre-atmospheric states for the observed meteoroids. All CAMS data are submitted regularly to the IAU Meteor Data Center, and are additionally hosted on the CAMS project website\footnote{\url{http://cams.seti.org/}} for open-access research. In this work, the meteor orbits we use are drawn from the 471,582-event catalog available through the IAU Meteor Data Center\footnote{\url{https://ceres.ta3.sk/iaumdcdb/}}.

\subsection{GMN}
Established in 2018, GMN\footnote{\url{https://globalmeteornetwork.org/}} has evolved into an extensive collaboration focused on continuous optical surveillance of meteors worldwide. Today, it involves more than 1\,000 video meteor cameras hosted by amateur astronomers and professionals in 45+ countries worldwide, each relying on CMOS cameras and Raspberry Pi hardware. Each camera station uses wide-field optics capable of detecting meteors down to about magnitude +6.0, recording at 25 frames per second. Median radiant precision stands at 0.47$^{\circ}$, with approximately 20\% of meteors observed from four or more stations achieving a precision of 0.32$^{\circ}$ in radiant accuracy \citep{vida2020estimating,vida2021global,vida2022computing}. Frequent calibrations are performed to maintain accuracy, using refined astrometric and photometric techniques that correct for local environmental and hardware variations. Once detections are confirmed, the network uploads results to a central server, often within hours, enabling real-time tracking of significant fireball events and potentially facilitating rapid meteorite recoveries. Overall, GMN exemplifies modern, collaborative meteor science, highlighting how open-source methods can significantly enhance global monitoring of near-Earth meteoroid activity \citep{vida2021global}.

\subsection{SonotaCo}
The SonotaCo Network is a distributed Japanese consortium of primarily amateur astronomers (with participation from public observatories) that has produced one of the largest homogeneous multi-station video meteor orbit datasets \citep{sonotaco2009meteor,sonotaco2016observation,masuzawa2021sonotaco}. Typical stations used Watec-class high-sensitivity monochrome cameras with fast ($\sim$f/0.8--f/1.2) wide-field lenses (tens of degrees field of view), recording NTSC-rate interlaced video. Astrometry, photometry, and orbit reduction are performed using the \texttt{UFOCapture}, \texttt{UFOAnalyzer}, and \texttt{UFOOrbit} software suite, with absolute timing synchronized via network time protocols (NTP) \citep{sonotaco2009meteor}. Published characterizations of the dataset indicate that detections are largely at visual magnitudes $\lesssim +2$. SonotaCo orbits have been extensively used as an independent reference sample for cross-survey meteor-shower confirmation and stream searches \citep[e.g.,][]{jenniskens2016verification}.

\subsection{Near-Earth asteroid sample}
All asteroid orbits were obtained from NASA’s JPL Horizons system\footnote{\url{https://ssd.jpl.nasa.gov/horizons/}}, which provides ephemerides and state vectors for every numbered and provisional small body. We retrieved the complete list of Near-Earth Objects (NEOs; 35\,012 at J2000) and then removed any object that satisfied the cometary-orbit criterion of \citet{tancredi2014criterion}. 

\section{Methods} \label{sec:methods}
The expected tidal signature within meteor/fireball catalogs is very small. In the NEA catalog, a low-$D$ excess became apparent only after the sample exceeded $\sim$20{,}000 objects \citep{jopek2020orbital,shober2025decoherence}. By analogy, a comparable number of sporadic, asteroidal meteors is likely required before an analogous signal could be detected robustly. Extending analyses to progressively fainter meteors increases raw statistics but also increases shower/comet contamination and strengthens the influence of nongravitational perturbations (e.g., Poynting--Robertson drag and solar radiation pressure), which can decorrelate orbits on shorter timescales. 

\subsection{Cometary Component Removal}
The cometary and shower components were removed using the following filtering steps: 
\begin{enumerate}
    \item removed all objects with $T_{J}<3$
    \item removed all bodies with $T_{J}>3$ found to be on cometary orbits deemed by the criterion set by \citet{tancredi2014criterion}
    \item removed all impacts where $D_{N}<0.2$ \citep{valsecchi1999meteoroid} or $D_{SH}<0.2$ \citep{southworth1963statistics} with an established or working meteor shower, as defined by the IAU Meteor Data Center
    \item removed all duplicate detections within CAMS 
\end{enumerate}

This strict filtering eliminated the vast majority of the raw meteor detections ($>$\,90\%). In addition, a small portion of the NEA data was removed for having either $T_J<3$ \citep{carusi1985long,levison1994long} or they fulfilled the ``cometary orbit'' criterion defined by \citet{tancredi2014criterion}, which has been shown to be a more nuanced way to identify possible cometary bodies compared to using only the Tisserand's parameter relative to Jupiter \citep{shober2024comparing}. However, we found it necessary to be certain that any signal was confidently asteroidal and unrelated to known non-tidal showers.

\subsection{Observational bias removal for perihelion distribution comparison}
Optical meteor and fireball detection systems are inherently more sensitive to faster meteoroids at the small end of the size distribution, resulting in an overrepresentation of high-velocity meteoroids in that regime. To correct this, we follow the approach of \citet{shober2025perihelion}, identifying the size range at which the observed size-frequency distribution begins to deviate from a power-law trend. All data below this cutoff are excluded to eliminate the portion most affected by bias. Although weighting schemes can be applied to account for such biases, our dataset's small signal of interest motivates the simpler, more robust approach of removing the biased size regime entirely. Note that this is performed solely to produce Figure~\ref{fig:g16_comparison} and Figure~\ref{fig:q_distro_normalized}, both of which require a debiased meteor distribution. 

The following cutoffs were placed: 
\begin{itemize}
    \item CAMS: $m_{v}<1.0$
    \item GMN: $m_{\infty}> 5\times10^{-4}$\,kg 
    \item FRIPON: $m_{v}<-6.0$
    \item EDMOND: $m_{v}<-0.6$
    \item EN: $m_{\infty}> 0.005$\,kg
    \item GFO: $m_{\infty}> 1.0$\,kg
\end{itemize}
where $m_{v}$ is the peak absolute magnitude of the meteor and $m_{\infty}$ is the mass of the meteoroid at the top of the atmosphere. It is also important to note that these sources use different methods (dynamic versus photometric) to estimate initial masses, which can introduce systematic differences when comparing masses.

\subsection{Search for the recent activity (within last 50\,kyr)}
Dynamically young meteoroid streams remain clustered in phase space only for a limited time before planetary perturbations and differential precession erase statistically separable structure, typically on $\sim$10--50\,kyr timescales \citep[e.g.,][]{pauls2005decoherence,shober2025decoherence}. A standard first diagnostic for such recent activity is an excess of very small pairwise dissimilarities relative to what is expected from random associations within the sporadic background. However, with large catalogs the number of unique pairs grows as $N(N-1)/2$, so even extremely small random-pair probabilities can produce non-negligible counts at low-$D$. A global excess in the cumulative pair counts, therefore, (i) requires an explicit null model to quantify false positives, and (ii) does not by itself identify where in observable phase space the excess originates. We therefore use two complementary analyses: (A) a global cumulative similarity distribution (CSD) significance test, and (B) a localized, bin-resolved excess-of-similarity map in $(U,\lambda_\odot)$ that attributes the global low-$D$ signal to specific regions of phase space.

\subsubsection{Global cumulative similarity distribution (CSD)} \label{meth:csd}
In this work, we use two complementary dissimilarity criteria. For comparisons in heliocentric orbital-element space, including the multi-dataset overview CSD in Figure~\ref{fig:combined_cumulative} and the NEA analysis in Figure~\ref{fig:stat_sig_nea} and Table~\ref{tab:nea_pairs}, we use the orbital dissimilarity metric $D_H$ \citep{jopek1993remarks}. Whereas as majority of the meteor similarity analysis is done in geocentric parameter space (Figures~\ref{fig:stat_sig} and \ref{fig:stat_sig_2d}). We use the geocentric dissimilarity criterion $D_N$ \citep{valsecchi1999meteoroid} which compares meteors in the four-parameter space $(U,\theta,\phi,\lambda_\odot)$. Here $U$ is the dimensionless geocentric speed scaled by the Earth's orbital speed, $U \equiv v_g / v_\oplus$ with $v_\oplus = 29.78~\mathrm{km\,s^{-1}}$, $\lambda_\odot$ is the solar longitude (degrees, treated as periodic on $[0,360)$), and $(\theta,\phi)$ are the usual apex-centered angles defined from the geocentric radiant and $\lambda_\odot$ following \citet{valsecchi1999meteoroid}. 

The $D$-values produced by dissimilarity functions, in essence, act as distance-like measures in orbital-element space (though, strictly speaking, not all satisfy the axioms of a metric; see section 2.1 of \citealp{kholshevnikov2016metrics}). To produce groups of interest, one must first propose a suitable clustering algorithm and then determine the tuning of the associated hyperparameters so that the groups are meaningful and statistically significant relative to the local background. Some popular, widely-used clustering algorithms include: single-neighbor linking \citep{lindblad1971stream}, group around a mean orbit \citep{sekanina1970statistical}, density mapping \citep{welch2001new}, and HDBSCAN/DBSCAN \citep{sugar2017meteor,pena2025meteoroid}. Estimating suitable D-thresholds (or ``D-criteria'') that are statistically significant has been the subject of numerous papers over the past decades. Many studies have proposed formulas to estimate thresholds depending on the D-function and the database size (e.g., \citealp{lindblad1971stream,jopek1999meteoroid,jopek2017probability}). For $D_{H}$ specifically, preliminary grouping has frequently used $0.05$–$0.10$, while $D_H\!<\!0.15$ and $<\!0.10$ are typical for long‑period and Jupiter‑family comet streams, respectively \citep{jenniskens2016established}. For a succinct reference, we point you towards the recent review of \citet{courtot2025orbit}, or previous notable works that discuss similarity functions \citep{jopek1993remarks,valsecchi1999meteoroid,jenniskens2008meteoroid,kholshevnikov2016metrics,pena2024statistical}. Fixed ``canonical'' limits, however, are sample- and error-model dependent: with $N_{GMN}\!=\!122{,}943$ meteors there are $>10^{9}$ unique pairs, so even a small random-pair probability at $D\!\sim\!0.05$ produces a few hundred thousand false-positives. For these reasons, rather than adopt a single, literature-based hard cut (e.g., $D_H\!\sim\!0.05$–$0.15$), we quantify significance against an empirical null built by randomly sampling an estimate for the local sporadic background and define ``candidate pairs'' where the observed cumulative pair counts exceed the expected stochastic-association frequency. 

For each catalog, we compute all unique pairwise values $D_{ij}$ and form the cumulative similarity distribution
\begin{equation}
N(<D) \equiv \sum_{i<j} \mathbf{1}(D_{ij}\le D),
\end{equation}
We define an ``excess'' as the observed $N(<D)$ exceeding the expectation from an empirical sporadic null which is estimated using a kernel density estimator (KDE) approach \citep{vida2017generating,shober2024generalizable}. The KDE is employed to approximate the background distribution of sporadic meteoroid orbits, and we draw $N_{\text{dataset}}$ synthetic samples from this distribution to gauge the level of random clustering for an $N_{\text{dataset}}$-sized dataset of impacts \citep{shober2025decoherence}. We use \texttt{scikit-learn}’s \texttt{KernelDensity} class \citep{pedregosa2011scikit} to generate random samples from a Gaussian kernel density estimator (KDE) fitted to our dataset. KDE has been previously shown to be an ideal method for creating synthetic sporadic samples while preserving crucial covariances between the parameters \citep{vida2017generating}. The sampling procedure begins by randomly choosing base points from the dataset, with each point having an equal chance of selection unless sample weights are provided. For each chosen base point $\mathbf{x}_i$, Gaussian noise is added to yield a new sample point $\mathbf{x}_i' = \mathbf{x}_i + \mathcal{N}(0, h^2)$. Here, $\mathcal{N}(0, h^2)$ denotes a random draw from a Gaussian distribution with zero mean and variance $h^2$, while $h$ is the bandwidth parameter controlling the width of the Gaussian kernel. Repeating this process for all base points yields samples that mirror the KDE-estimated probability density function. 

Our previous work employed a single scalar bandwidth $h$ after Z-score standardization to investigate how smoothing affects false-positive rates in a controlled manner \citep{shober2024generalizable}. A single $h$ imposes spherical smoothing in the five-dimensional space; however, the sporadic population is strongly anisotropic—exhibiting narrow structure in $(q,e,i)$ (e.g., the steep rise toward $q\simeq 1~\mathrm{au}$) and broad, wrapped structure in $(\omega,\Omega)$. With a scalar $h$ one must either over-smooth the narrow $(q,e,i)$ ridges (reducing local density and under-predicting random close pairs at small $D$, which inflates apparent clustering significance) or under-smooth the angular coordinates (increasing random-pair counts and reducing sensitivity). Following standard practice for multivariate KDE with differing coordinate curvatures \citep[e.g.,][]{hastie2009elements,silverman2018density}, and consistent with the diagonal-bandwidth approach advocated for meteoroid spaces by \citet{vida2017generating}, we therefore adopt a diagonal bandwidth matrix. We fit a multivariate Gaussian KDE to $(q,e,i,cos\omega,\sin\omega,\cos\Omega,\sin\Omega)$ to estimate the sporadic null background when using $D_H$. Otherwise, when using geocentric parameters, we fit a 6D Gaussian KDE in $(U,\mathrm{atanh}(\cos\theta),\cos\phi,\sin\phi,\cos\lambda_\odot,\sin\lambda_\odot)$, which treats the angular coordinates as periodic via sine/cosine embedding while preserving covariances with $U$ and $\theta$. 

Bandwidth selection follows the improved Sheather--Jones method, with minor adjustments to ensure that broad features in the distributions are not over-smoothed. Explicitly, initial Sheather--Jones bandwidths are applied, and the KDE 3-$\sigma$ region is estimated (blue regions in Figures~\ref{fig:stat_sig_nea}-\ref{fig:stat_sig}). The bandwidth parameters are then equally increased or decreased until the KDE 3-$\sigma$ region corresponds well to the observed CSD at large D-values. At larger D-values, the CSD deviates from a power-law due to overall population distribution, i.e., unrelated to recent clustering/activity. This adjustment was small in every case. From the null ensemble, we compute the mean and standard deviation of $N(<D)$ at each $D$ threshold, and we evaluate the observed curve against the $+3\sigma$ upper envelope. Where available, we additionally propagate reported observational uncertainties assuming Gaussian errors; the corresponding $3\sigma$ band is then used as a conservative diagnostic of sensitivity to measurement uncertainty.

\subsubsection{Localized pair-excess mapping in $(U,\lambda_\odot)$}\label{meth:2d_csd}
A global CSD excess does not indicate which region(s) of phase space generate the low-$D_{\mathrm{N}}$ surplus; we localize the signal using a bin-resolved, \emph{pair-conditioned} midpoint-count statistic in $(U,\lambda_\odot)$. We discretize $(U,\lambda_\odot)$ into an $N_U\times N_{\lambda}$ grid with equal-width bins spanning the observed ranges, treating $\lambda_\odot$ as periodic on $[0,360)$.
For a chosen dissimilarity limit $D_{\rm lim}$ (here $D_{\rm lim}=0.015$), we compute the observed binwise pair-count field
\begin{equation}
N^{\rm obs}_{mn}(D_{\rm lim}) \equiv \sum_{i<j}
\mathbf{1}(D_{ij}\le D_{\rm lim})\,
\mathbf{1}\!\left[(U_{ij}^{\rm mid},\lambda_{\odot,ij}^{\rm mid})\in \mathcal{B}_{mn}\right],
\end{equation}
where $\mathcal{B}_{mn}$ denotes bin $(m,n)$, $U_{ij}^{\rm mid}=(U_i+U_j)/2$, and the circular midpoint in solar longitude is
\begin{equation}
\lambda_{\odot,ij}^{\rm mid} \equiv 
\operatorname{atan2}\!\left(\sin\lambda_{\odot,i}+\sin\lambda_{\odot,j},\,
\cos\lambda_{\odot,i}+\cos\lambda_{\odot,j}\right)\bmod 360^\circ,
\end{equation}
with trigonometric functions evaluated on angles in radians and the result mapped to degrees on $[0,360)$. (The antipodal case is measure-zero in practice; when it occurs we assign $\lambda_{\odot,ij}^{\rm mid}=\lambda_{\odot,i}$.)

To assess significance, we generate $K$ KDE-drawn synthetic catalogs from the same sporadic null used for the global CSD and compute the corresponding null pair-count fields $N^{(k)}_{mn}(D_{\rm lim})$. From the null ensemble we estimate, independently in each bin,
\begin{equation}
\mu_{mn} \equiv \frac{1}{K}\sum_{k=1}^{K} N^{(k)}_{mn},\qquad
\sigma_{mn} \equiv \sqrt{\frac{1}{K-1}\sum_{k=1}^{K}\left(N^{(k)}_{mn}-\mu_{mn}\right)^2},
\end{equation}
and define a per-bin $z$-score $z_{mn}\equiv (N^{\rm obs}_{mn}-\mu_{mn})/\sigma_{mn}$.
We flag a bin as significant if $N^{\rm obs}_{mn}>\mu_{mn}+k\sigma_{mn}$ with $k=3$. As the $z_{mn}$ values are evaluated over a scan of $m=N_U N_{\lambda}$ bins, they are local statistics. When quoting a global false-alarm probability, we convert the maximum observed $z$ to a one-sided $p_{\rm local}$ under a standard-normal tail and apply a Dunn--\v{S}id\'ak family-wise adjustment $p_{\rm global}=1-(1-p_{\rm local})^{m}$ \citep{dunn1961multiple,sidak1967rectangular}.

For visualization we define the gated surplus map
\begin{equation}
S_{mn} \equiv (N^{\rm obs}_{mn}-\mu_{mn})\,\mathbf{1}(z_{mn}\ge k),
\end{equation}
where $\sigma_{mn}$ includes a small numerical floor to avoid divide-by-zero in sparsely populated bins.

Monte Carlo estimates of $(\mu_{mn},\sigma_{mn})$ can be noisy in bins with low null occupancy, so when adaptive sampling is enabled we monitor convergence of the exceedance threshold $T_{mn}\equiv \mu_{mn}+k\sigma_{mn}$ as additional null draws are accumulated. In particular, for bins currently satisfying $N^{\rm obs}_{mn}>\mu_{mn}+k\sigma_{mn}$ we estimate the standard error of the threshold via
\begin{equation}
\mathrm{SE}(\mu_{mn})=\frac{\sigma_{mn}}{\sqrt{K}},\qquad
\mathrm{SE}(\sigma_{mn})\approx\frac{\sigma_{mn}}{\sqrt{2(K-1)}},\qquad
\mathrm{SE}(T_{mn})\approx\sqrt{\mathrm{SE}(\mu_{mn})^2+k^2\mathrm{SE}(\sigma_{mn})^2},
\end{equation}
and require a robust ``gap'' condition
\begin{equation}
N^{\rm obs}_{mn}-T_{mn} \ge g\,\mathrm{SE}(T_{mn}),
\label{eq:min_SE}
\end{equation}
with $g=3$, so that bins flagged as significant are not artifacts of Monte Carlo noise in the null threshold. No bins are highlighted that do not satisfy eq.~\ref{eq:min_SE}. The minimum number of KDE-drawn synthetic catalogs $K$ was set to 200, and the maximum was set to 5000. The four meteor catalogues required 400, 1500, 2200, and 3000 synthetic draws, respectively, for GMN, CAMS, SonotaCo, and EDMOND to satisfy the condition of eq.~\ref{eq:min_SE} (the variation is almost entirely due to the specific database's total size).

Finally, to interpret localized radiant-space excesses in orbital-element space, we identify meteors that participate in at least one significant-bin pair at the adopted $D_{\rm lim}$. For each significant bin $(m,n)$ we collect the unique set of contributing meteors and distribute that bin's surplus $S_{mn}$ uniformly across them; summing over significant bins yields a per-meteor weight proportional to the localized pair surplus. We then reproject the weighted meteors into orbital $(q,i)$ space by forming a weighted 2D histogram over the observed $(q,i)$ ranges. 


\subsection{Search for long-term tidal signature (last several millions of years)}
To search for long-term tidal disruption signatures, we follow the methodology of \citet{granvik2024tidal}, who identified statistically significant excesses in the perihelion distribution of the NEAs observed by Catalina Sky Survey during 2005-2012 near the orbit of the Earth and the orbit of Venus (Figure 1 in \citealt{granvik2024tidal}). Their approach compares the observed distribution to a synthetic population generated by the debiased NEO model of \citet{granvik2016super}, which does not incorporate tidal disruption effects. Here, we have also similarly compared the q-distribution of 18\,012 meteors and fireballs observed by eight different sources (see Fig.~\ref{fig:g16_comparison}) to the debiased NEO model of \citet{granvik2016super}. Specifically, we test for an excess of orbits near q$\approx$0.7\,au (Venus) and q$\approx$1\,au, consistent with expectations for tidally released debris. This approach is complementary to recent work by \citet{chow2025decameter}, who applied it to Electron-Multiplied Charge-Coupled Device (EMCCD) camera observations made by the Canadian Automated Meteor Observatory and found no strong tidal signal. This perihelion analysis uses the separate observation-debiased multi-source sample (N=18{,}012) rather than the full 235{,}271-orbit video-meteor sample used in the localized excess mapping.

\section{Results and Discussion} \label{sec:results}

\begin{figure}[ht!]
\plotone{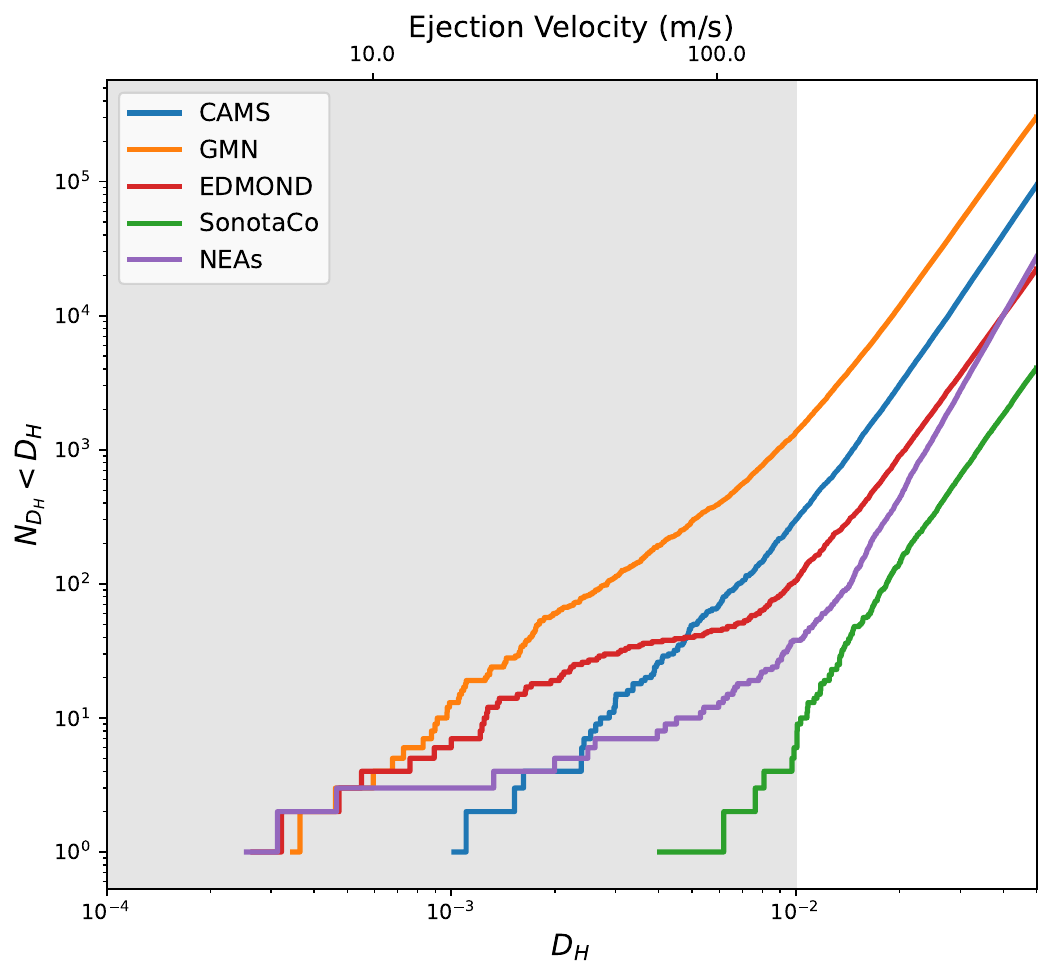}
\caption{Cumulative similarity distribution of all possible pair combinations within the shower-removed asteroidal subsets of the GMN, CAMS, EDMOND, and SonotaCo video-meteor catalogs (N=235{,}271), together with the NEA sample (N=35{,}012), using the orbital dissimilarity metric $D_H$ \citep{jopek1993remarks}.
\label{fig:combined_cumulative}}
\end{figure}

In an ideal self-similar ensemble, the CSD should follow a power-law at small values until it runs out of statistics. However, at small $D_H$ values (and $D_N$), we find that the CSD curve deviates from linearity, indicative of an excess of similarity being recorded (Figure~\ref{fig:combined_cumulative}). The sources that have a large enough dataset to extend below $D_H=10^{-2}$ (CAMS, GMN, EDMOND, and the NEA catalog) all have an excess of similarity below this limit; see how the curves diverge from linear. A similar break was reported by \citet{shober2025decoherence}, but at a slightly higher similarity threshold ($D_{H}\sim0.03$). The difference is traceable to cometary fragments, most notably the chain of pieces from comet 73P/Schwassmann–Wachmann 3, which were included in that earlier NEO sample. When we remove the near-Earth comets and apply the \citet{tancredi2014criterion} criterion to remove all bodies on comet-like orbits, the NEA cumulative curve slightly shifts: the comet-driven bump at $D_{H}\sim0.03$ disappears, and the residual distribution shows the $10^{-2}$ kink we see in the asteroidal meteor data. 

To provide concrete examples of the extreme low-$D_{\mathrm H}$ tail in the NEA sample, Table~\ref{tab:nea_pairs} lists the ten tightest NEA--NEA pairs with $D_{\mathrm H}<0.005$. Although the population-level perihelion excess near $q\simeq a_{Earth}$ (and $q\simeq a_{Venus}$) relative to steady-state debiased NEO models is consistent with encounter-driven tidal disruption \citep{granvik2024tidal}, no ``smoking-gun'' tidal-disruption family emerges from the pair list. In particular, two low-MOID pairs in Table~\ref{tab:nea_pairs} (2003~EW59--2023~RP and 2019~UN13--2023~VA) remain geometrically compatible with very recent encounter-driven fragmentation, but their high condition codes preclude a definitive interpretation at present; nevertheless, their orbital uncertainties are still typically small compared to those of most meteor-derived orbits, so they remain viable tidal-disruption candidates pending orbit refinement and physical follow-up. Conversely, several of the most robustly characterized NEA-pair cases in the literature are consistent with non-tidal formation channels (e.g., YORP-driven rotational fission and/or binary dissociation) \citep[e.g.,][]{moskovitz2019common}, emphasizing that multiple mechanisms can populate the low-D regime. Relaxing the similarity threshold to larger $D_{\mathrm H}$ would certainly increase the number of candidates, but at the cost of a rapidly rising false-positive fraction \citep{shober2025decoherence}; therefore, the most direct path toward the first unambiguous tidal-disruption family is improved discovery statistics, orbit quality, and physical characterization for small, Earth-like NEAs, which will be enabled by next-generation surveys and follow-up programs (e.g., Rubin/LSST and NEO Surveyor; \citealt{ivezic2019lsst,mainzer2023near}).

\begin{table}[h]
    \centering
    \caption{NEA pairs with $D_H<0.005$}
    \label{tab:nea_pairs}
    {\nolinenumbers
    \footnotesize
    \begin{tabular}{llp{4.0cm}llrrl}
        \toprule
        NEA 1 & NEA 2 & Interpretation(s) & Previous Works & Spectra & MOID (LD) & $D_H$ & Orb. quality \\
        \midrule
        2019 PR2  & 2019 QR6   & Very recent genetic pair; possibly rot. fission; comet-like nongrav forces likely post-separation” & \citealt{fatka2022recent} & D & 91.8 & 0.00025 & 0/0 \\
        2012 QG8  & 2024 GE1   &  & \citealt{pereira2025photometric} & Cg \& ? & 89.1 & 0.00031 & 0/7 \\
        2022 MY3  & 2022 OP8   &  &  &  & 72.5 & 0.00046 & 1/7 \\
        2003 XE   & 2023 YD2   &  &  &  & 52.3 & 0.00133 & 0/7 \\
        2017 SN16 & 2018 RY7   & YORP spin-up and/or binary dissociation; separation age <10 kyr. & \citealt{moskovitz2019common} & V & 36.4 & 0.00200 & 1/2 \\
        2010 VL65 & 2021 UA12  & Not a pair. Misidentified single-object due to nongrav & \citealt{taylor2024strong} &  & 2.5  & 0.00249 & 6/6 \\
        2015 EE7  & 2015 FP124 & YORP spin-up and/or binary dissociation. & \citealt{moskovitz2019common} & S-complex & 25.4 & 0.00262 & 0/8 \\
        2003 EW59 & 2023 RP    &  &  &  & 5.2  & 0.00369 & 8/7 \\
        2005 TF   & 2024 WS19  &  & \citealt{ieva2020extended} & S-complex \& ? & 35.9 & 0.00418 & 0/5 \\
        2019 UN13 & 2023 VA    &  &  &  & 0.045 & 0.00450 & 7/6 \\
        \bottomrule
    \end{tabular}
    }
    \tablecomments{Based on the background null estimate (blue line Figure~\ref{fig:stat_sig_nea}), about 50\% of these pairs are false-positives. The orbit quality values (cond$_1$/cond$_2$) correspond to the NASA HORIZONS ``condition\_code'' metric used to measure observation quality. The value ranges from 0 to 9, where lower values indicate a more precisely determined orbit.}
\end{table}

\begin{figure}[ht!]
\centering
\plotone{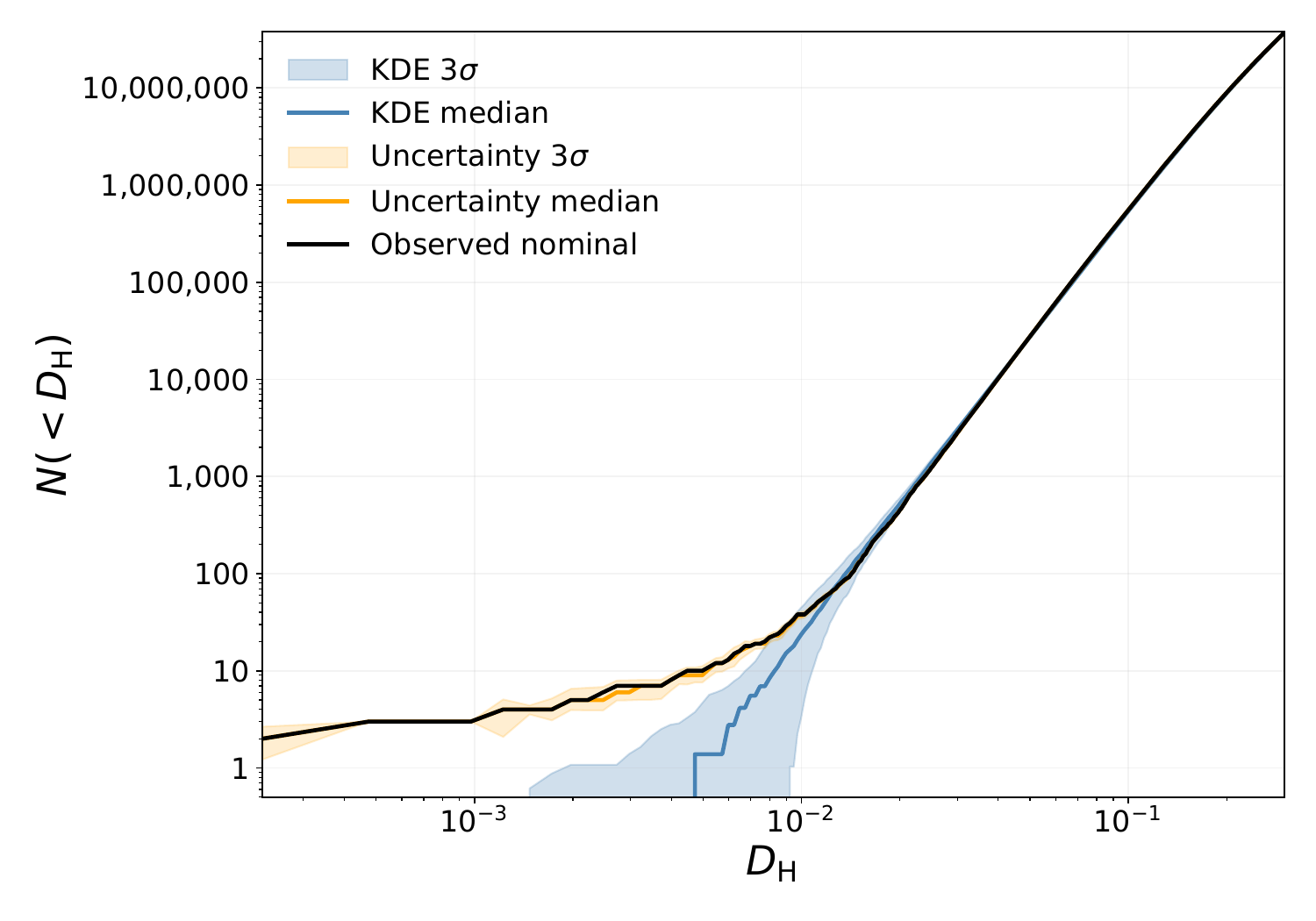}
\caption{
Cumulative number of NEA--NEA pairs with $D_H$ \citep{jopek1993remarks} values below a limiting threshold. The black line uses the nominal orbital parameters, while the yellow line and shaded region include observational uncertainties. The blue-shaded area denotes the $3\sigma$ range expected from random associations based on KDE-drawn synthetic samples.
}
\label{fig:stat_sig_nea}
\end{figure}

\begin{figure*}[ht!]
\centering
\gridline{
  \fig{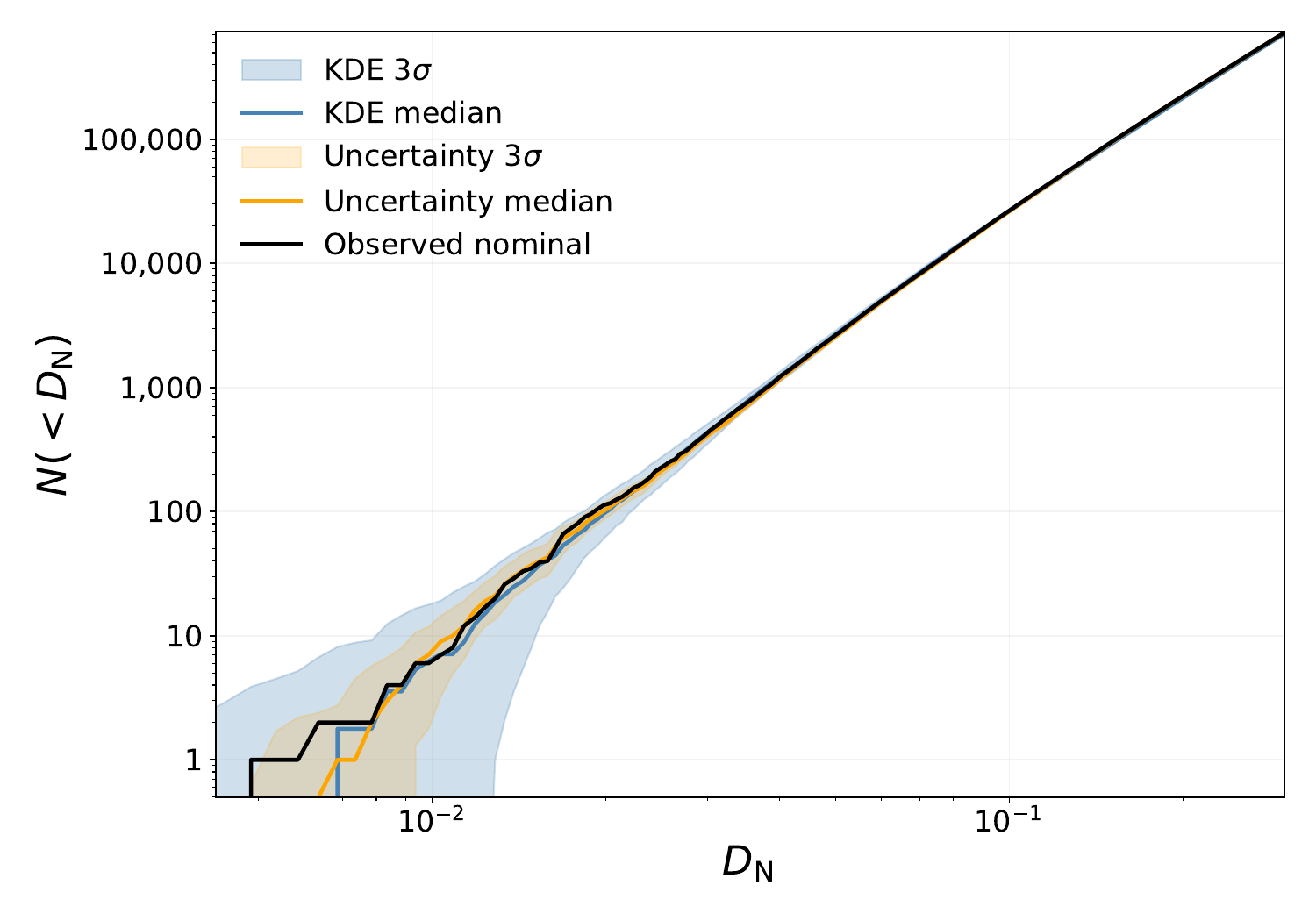}{0.45\textwidth}{(a) Sonotaco}
  \fig{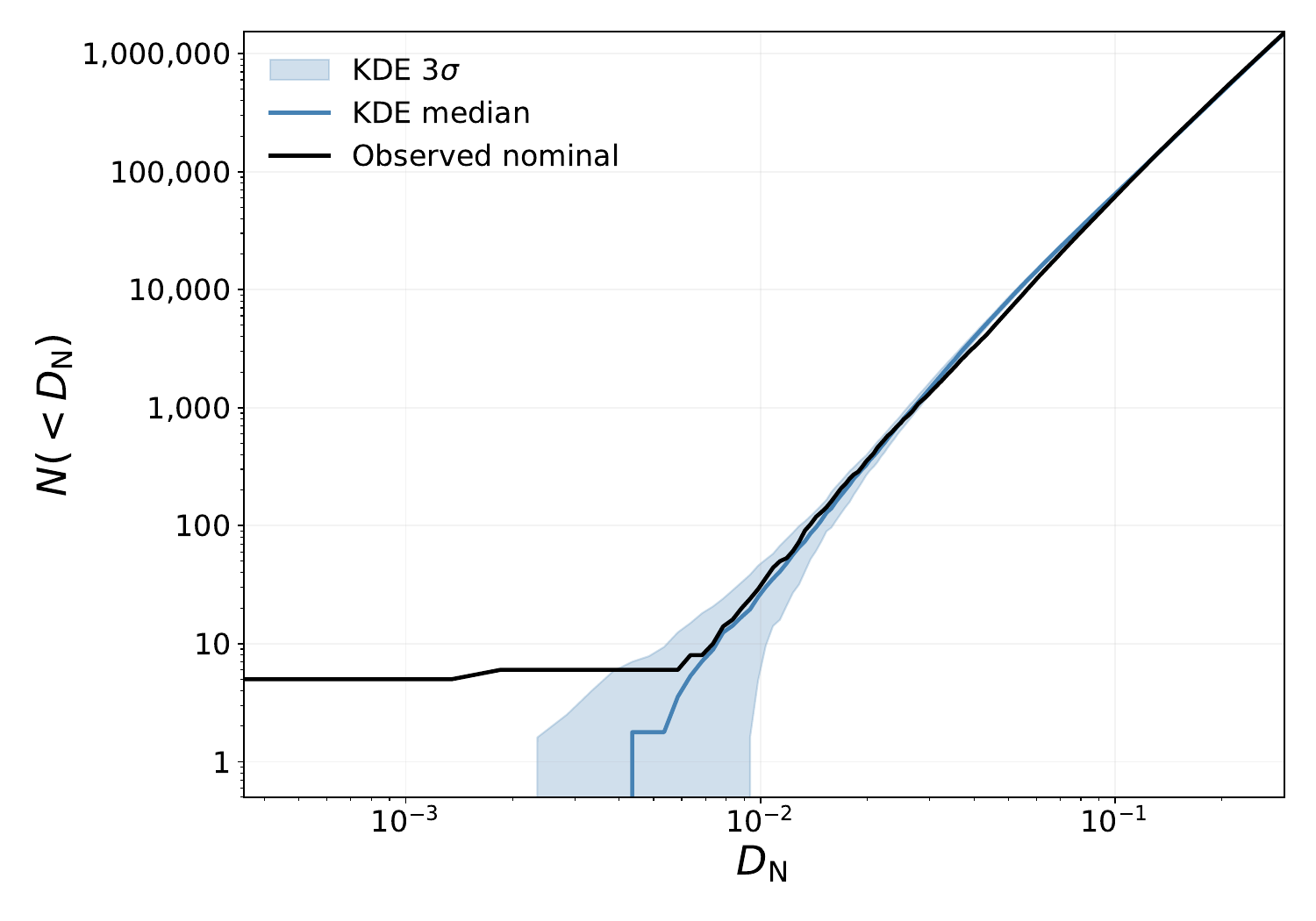}{0.45\textwidth}{(b) EDMOND}
}
\gridline{
  \fig{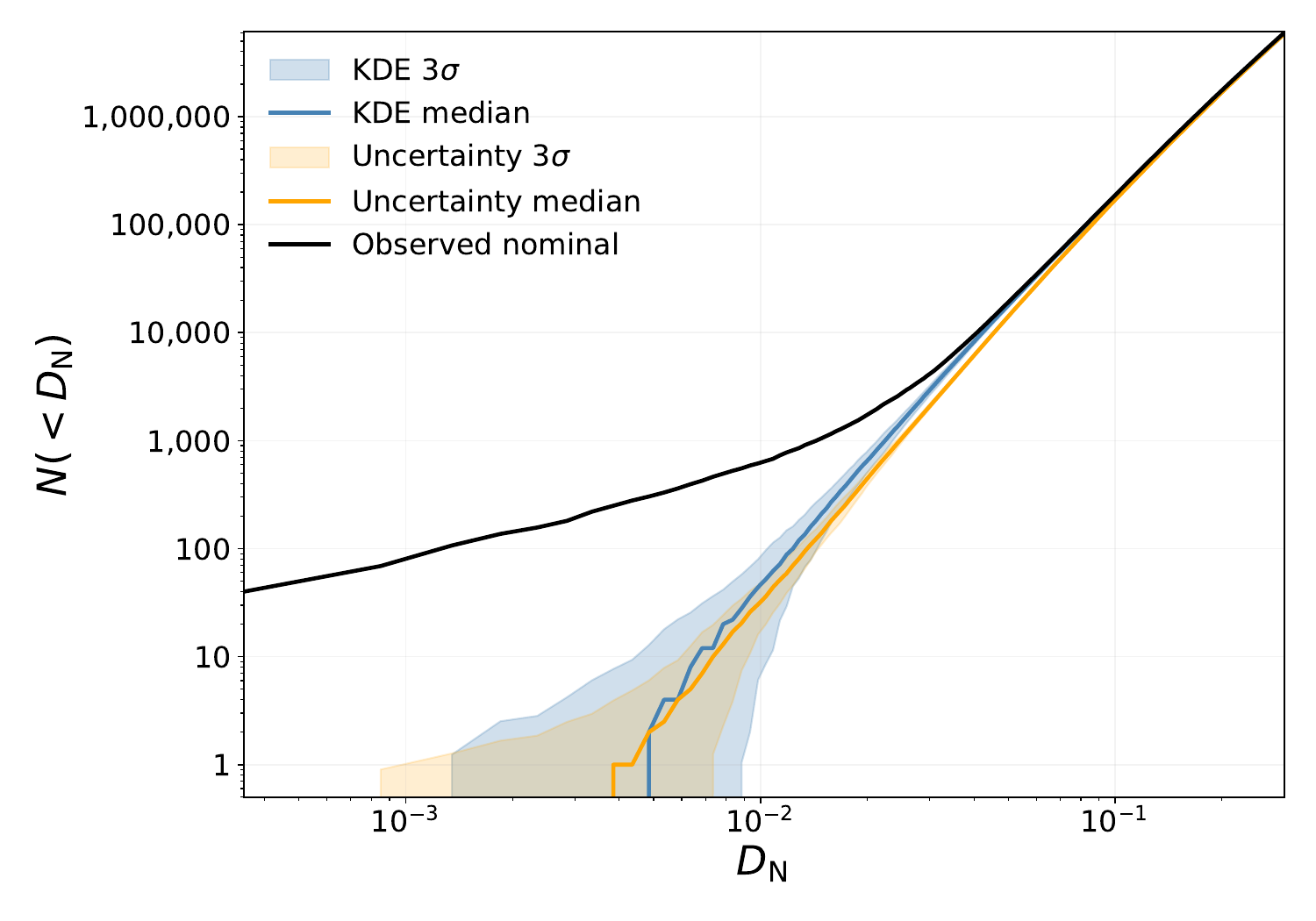}{0.45\textwidth}{(c) CAMS}
  \fig{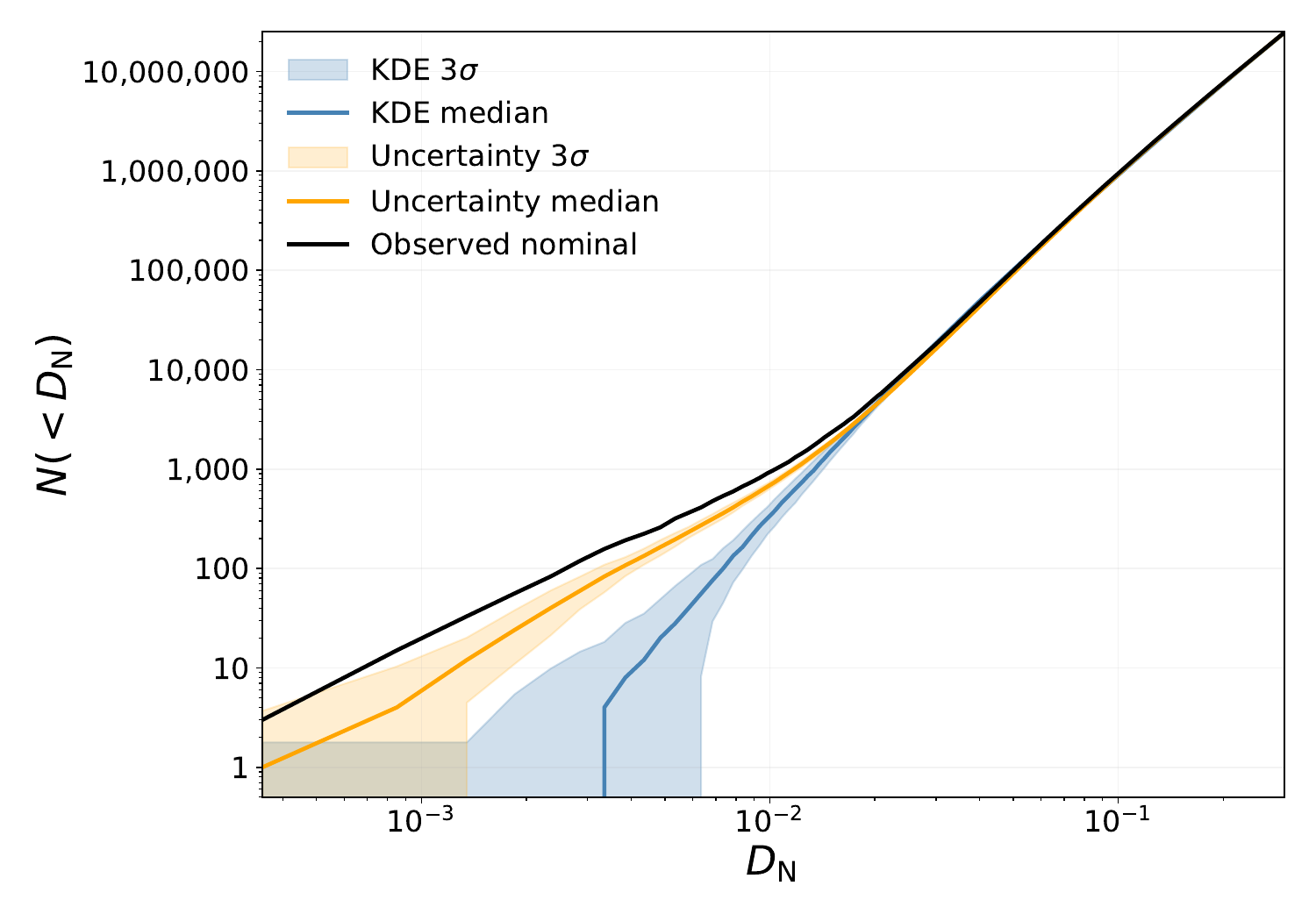}{0.45\textwidth}{(d) GMN}
}
\caption{Number of (a) SonotaCo--SonotaCo, (b) EDMOND--EDMOND, (c) CAMS--CAMS, and (d) GMN--GMN pairs, $N_D$, with $D_N$ values below a limiting threshold compared to the predicted number from random associations (blue line with a $3\sigma$ region) based on KDE-drawn synthetic samples simulating the sporadic background population. The black line uses the nominal observed geocentric parameters, while the yellow line and shaded region include observational uncertainties. The cumulative number of low-$D_N$ pairs is slightly in excess for three datasets, but only GMN shows an excess that remains above the uncertainty band.
\label{fig:stat_sig}}
\end{figure*}

Figures~\ref{fig:stat_sig_nea} and \ref{fig:stat_sig} quantify the significance of the first break in the CSD slope for the NEAs and meteor datasets, respectively. We compare the observed pair counts with the expectations from a kernel-density model of the sporadic asteroidal population, constructed using the method of \citet{shober2024generalizable}. In every dataset where the CSD curve crosses $D_H=10^{-2}$, the observed number of pairs below that threshold exceeds the stochastic prediction by $>3\sigma$. The only dataset to not display this change in slope is SonotaCo; however, this is because the sporadic subset is 3-5$\times$ smaller than the other databases and there is not good statistics for $D_H<10^{-2}$. This cumulative excess among meteor datasets is, nonetheless, suspect, as the similarity completely disappears for CAMS (and possibly EDMOND, but no formal uncertainties published)  when nominal observational uncertainties are included (yellow region in Figures~\ref{fig:stat_sig_nea} and \ref{fig:stat_sig}).

CAMS exhibits an extensive excess at low-$D_N$ (Figure~\ref{fig:stat_sig}). CAMS metadata includes an internal duplicate flag; we remove all events carrying this flag prior to pair counting. These events are consistent with previous observations of larger ``meteor cluster’’ events, where several to tens of meteors are observed within seconds of each other \citep{koten2017clusterseptember,capek2022ejection,koten2024clusterproperties,koten2025clustervery,ashimbekova2025towards}. \citet{ashimbekova2025towards} recently identified 16 new cluster candidates, each involving 4–7 meteoroid fragments, that met strict statistical significance criteria. Detailed analyses of these events predict separation velocities of less than 1\,m/s, which typically occur within days to a week of atmospheric impact. The most likely formation scenario proposed is thermal stress fracturing. As an additional cross-check, we applied a conservative time--space proximity filter that removes meteors observed within 5~s of each other and with reconstructed begin-point separations $<10$~km (thereby removing both potential unflagged duplicates or ``meteor clusters''). In this sensitivity test, the CAMS cumulative excess similarity at low-$D_H$ completely disappears, while only a small fraction of the other networks' excesses are removed. This supports the interpretation that the CAMS low-$D$ excess is dominated by a mixture of unresolved duplicates and meteor-clustering episodes consistent with fragmentation shortly before atmospheric entry \citep{koten2017clusterseptember,capek2022ejection,koten2024clusterproperties,koten2025clustervery,ashimbekova2025towards}.

Primary culprits of the ``kink’’ at $D_H\lesssim10^{-2}$ could include tidal disruption, thermal fracturing, along with meteoroid impacts, and spin-up disruption. However, disentangling these mechanisms is complicated by the fact that their ejection-velocity ranges are normally indistinguishable in meteor datasets. Slow tidal disruption events and thermal cracking both yield $v_{\rm ejection}\approx0.1$–1 m s$^{-1}$ \citep{zhang2020tidal,capek2022ejection}. Meanwhile, meteoroid impacts, for example, have been observed on asteroid Bennu to eject particles at speeds of a few meters per second \citep{lauretta2019episodes,bottke2020meteoroid}. Multi-station video and fireball networks typically deliver speed uncertainties of 50–150 m s$^{-1}$, ideally. Such errors blur orbital-similarity metrics to $D_H\gtrsim10^{-3}$--$10^{-2}$, effectively erasing the ultra-low-$D_H$ regime unless the meteors are recorded within seconds or minutes of one another. While the velocity range is an order of magnitude larger than that of tidal disruptions or thermal fragmentations, the inherent velocity uncertainties are significantly larger. Improving velocity precision can help with this; however, ultimately, we either require additional observations (e.g., spectra) or need to examine the orbital regimes in which this excess similarity appears. To dissect this excess of similarity found within the asteroidal and meteor datasets further, we generalized the CSD methodology to a 2D phase space of $(U,\lambda_\odot)$ (see section~\ref{meth:2d_csd}).

\subsection{Confirmation of a recent, diffuse Virginid stream}\label{subsec:new_svir_stream}

Across all four networks, Figure~\ref{fig:stat_sig_2d} does not show a clear excess concentrated in the tidal-disruption-like region ($q\sim 1$~au and $i<20^\circ$). Instead, the dominant and most coherent feature in the orbit-space projection is the GMN excess at low $q$ and low $i$, which originates from the most significant GMN 2D excess bin at $U\in[0.967,1.061]$ and $\lambda_\odot\in[0^\circ,18^\circ]$ with $N_{\rm obs}=135$ pairs versus $\langle N_{\rm null}\rangle=38.94$ (corresponding to $z=6.32$ and a gated surplus of $N_{\rm obs}-\langle N_{\rm null}\rangle=96.06$ pairs). This GMN-excess corresponds to a one-sided $p_{\rm local}=1.3\times10^{-10}$ and a Dunn--\v{S}id\'ak corrected $p_{\rm global}=5.2\times10^{-8}$. The same low-$q$/low-$i$ bin also shows a very modest (+0.5-1.5 pairs) excess in the SonotaCo and EDMOND datasets. The only dataset that does not show an excess in any way is CAMS; however, its primarily northern hemisphere observations ($\sim$94\%) in Database v3.0:2010-2016 likely explain this. This very significant cluster maps to a coherent radiant region within the broader Virginid sky area and currently does not correspond to any known established or working-group meteor shower using orbital \citep{southworth1963statistics,jopek1993remarks} or geocentric \citep{valsecchi1999meteoroid} similarity functions. Including the MDC list of removed showers, we found that activity in this region was previously submitted to the MDC as the candidate shower \textit{87 Virginids} (IAU \#01185; code ESV), which is currently listed as `removed' in the MDC database (see \citet{jenniskens2020removing} and \citet{hajdukova2023modification} for details on why showers are removed). Initial comparisons did not identify this association with the removed shower, as the orbital elements are not correctly cataloged. The only reference to the ``87 Virginids'' appears in \citet{jenniskens2023atlas} and in Table 1 in \citet{jenniskens2024properties}; however, this study is the first to establish the statistical significance of the cluster. The stream has been added by the IAU-MDC to the working list under the new provisional name M2026-A1.

We isolated stream members via density-based clustering. Applying DBSCAN ($\varepsilon_{D_N}=0.03$, $\mathrm{min_{samples}}=2$; \citealt{ester1996density}, consistent with previous work such as \citealt{sugar2017meteor}) yields a connected cluster of $N=282$ meteors. The cluster is strongly GMN-dominated ($243/282=86.2\%$), with additional members drawn from SonotaCo ($19$), CAMS ($10$), and EDMOND ($10$). A second DBSCAN search was also done, including 41\,176 theoretical NEA radiants calculated using the ``W-method'' as described in \citet{neslusan1998computer}, but none were associated with this shower. The shower activity spans $\lambda_{\odot}=359.22^{\circ}$ to $19.74^{\circ}$ (J2000), and appears to be annual based on detections ranging from 2006-2025. The radiant, velocity, and radiant drift can be found in Table~\ref{tab:mdc_mean_svir_activity}. Meanwhile, the orbital information can be found in Table~\ref{tab:mdc_mean_svir_orbit}. The median heliocentric orbit of the cluster indicates a highly evolved, asteroidal source detached from the main-belt and reaching very close to the Sun ($a=1.29\pm 0.10~\mathrm{au}$, $q=0.22\pm 0.01~\mathrm{au}$, $i=12.3^{\circ}\pm 1.8^{\circ}$). 

Given its very small perihelion distance and highly dispersed radiant, this newly identified Virginid-region stream is consistent with recent, thermally driven activity (a ``rock-comet'' style mass-loss pathway) in which a small, weak (potentially carbonaceous) NEA shed debris through near-Sun thermo-mechanical processing rather than ice sublimation. At $q\simeq0.22$~au, a plausible contributor could be thermal fracture/fatigue \citep{delbo2014thermal,molaro2020situ} along with dehydration cracking and mineral decomposition-driven degassing \citep{jewitt2012active,maclennan2024thermal,tsirvouils2026instantaneous}, and subsequent radiation-pressure sweeping, as suggested by perihelion activity of (3200)~Phaethon \citep{jewitt2010activity,jewitt2013dust}. In the case of Phaethon, proposed near-perihelion gas production arises from thermal decomposition of carbonates, Fe-sulfides, and phyllosilicates, releasing CO$_2$, S$_2$, and H$_2$O \citep{maclennan2024thermal}.

For the GMN-dominated cluster, we additionally estimated meteoroid robustness using the end-height PE parameter and the pressure-resistance factor \(Pf\) (a strength proxy based on the maximum dynamic pressure scaled by mass, speed, and entry geometry; \citealt{borovivcka2022_two}). The distribution is narrowly centered at \(\mathrm{median}\,PE\simeq-5.1\) (predominantly PE type~II) and \(\mathrm{median}\,Pf\simeq0.17\) (predominantly Pf-III), implying ``shower-like'' fragmentation resistance that is substantially lower than the strongest asteroidal-stream endmember (e.g., Geminids/$\eta$~Virginids in the \citealt{borovivcka2022_two} scheme) but also not as weak as the most fragile cometary cases (Pf-IV/V). This moderate \(Pf\) is therefore compatible with a very recent, thermally driven, near-Sun degradation pathway, where repeated thermo-mechanical damage and devolatilization/decomposition have not yet broken down the meteoroids significantly. 

This diffuse, low-$q$ stream is also an attractive target for identifying associated fragments or parent candidates with next-generation surveys. NASA's NEO Surveyor, planned for launch no earlier than September 2027, will conduct a wide-field mid-infrared survey from Sun--Earth L1, with sensitivity to low-albedo objects and access to small solar elongations, providing diameter constraints (and, with optical follow-up, albedos) that are difficult to obtain from reflected light alone \citep{mainzer2023near}. 

\begin{deluxetable*}{lccrrrrrrrrr}
\tabletypesize{\scriptsize}
\tablewidth{0pt}
\setlength{\tabcolsep}{4pt}
\renewcommand{\arraystretch}{1.15}
\tablecaption{Provisional southern Virginid-region stream (IAU-MDC code: M2026-A1): activity, geocentric radiant, radiant drift, and geocentric speed, reported by source network and as a combined solution. \label{tab:mdc_mean_svir_activity}}
\tablehead{
\colhead{Source} &
\colhead{$N$} &
\colhead{Years} &
\colhead{$\lambda_{\odot,b}$} &
\colhead{$\lambda_{\odot,\max}$} &
\colhead{$\lambda_{\odot,e}$} &
\colhead{$\alpha_{g}$} &
\colhead{$\delta_{g}$} &
\colhead{$d\alpha_g/dt$} &
\colhead{$d\delta_g/dt$} &
\colhead{$V_{g}$} \\
\colhead{} &
\colhead{} &
\colhead{} &
\colhead{(deg)} &
\colhead{(deg)} &
\colhead{(deg)} &
\colhead{(deg)} &
\colhead{(deg)} &
\colhead{(deg/day)} &
\colhead{(deg/day)} &
\colhead{(km~s$^{-1}$)}
}
\startdata
GMN      & 243 & 2020--2025      & 0.2   & 11.8 & 19.7 & $209.0\pm3.4$ & $-20.3\pm2.0$ & $+0.86$ & $-0.42$ & $29.8\pm0.9$ \\
CAMS     & 10  & 2012/2015/2016  & 7.2   & 15.0 & 17.6 & $210.8\pm3.6$ & $-20.9\pm2.0$ & $+0.95$ & $-0.33$ & $29.9\pm0.8$ \\
EDMOND   & 10  & 2006-2020       & 359.2 & 15.4 & 17.4 & $212.1\pm4.5$ & $-21.2\pm1.7$ & $+0.80$ & $-0.22$ & $30.3\pm1.3$ \\
SonotaCo & 19  & 2009-2023       & 7.2   & 10.2 & 18.3 & $208.4\pm3.2$ & $-19.3\pm1.6$ & $+0.92$ & $-0.32$ & $29.6\pm0.9$ \\
Combined & 282 & annual          & 359.2 & 11.9  & 19.7 & $209.1\pm3.5$ & $-20.3\pm1.9$ & $+0.86$ & $-0.41$ & $29.8\pm0.9$ \\
\enddata
\tablecomments{
All quantities are J2000. $\lambda_{\odot,b}$ and $\lambda_{\odot,e}$ are the minimum and maximum of the unwrapped solar-longitude distribution within each subset. $\lambda_{\odot,\max}$ is the median of the $\lambda_{\odot}$ distribution. $\sigma_{\alpha_{g}}$, $\sigma_{\delta_{g}}$, $\sigma_{V_g}$ are the $1\sigma$ dispersion in each subset (reported to provide a checkable spread rather than a formal uncertainty on the median).
}
\end{deluxetable*}

\begin{deluxetable*}{lcrrrrrrr}
\tabletypesize{\scriptsize}
\tablewidth{0pt}
\setlength{\tabcolsep}{4pt}
\renewcommand{\arraystretch}{1.15}
\tablecaption{Heliocentric orbital elements for southern Virginid-region stream (IAU-MDC code: M2026-A1), reported by source network and as a combined solution.\label{tab:mdc_mean_svir_orbit}}
\tablehead{
\colhead{Source} &
\colhead{$N$} &
\colhead{$a$} &
\colhead{$q$} &
\colhead{$e$} &
\colhead{$i$} &
\colhead{$\omega$} &
\colhead{$\Omega$} &
\colhead{$T_J$} \\
\colhead{} &
\colhead{} &
\colhead{(au)} &
\colhead{(au)} &
\colhead{} &
\colhead{(deg)} &
\colhead{(deg)} &
\colhead{(deg)} &
\colhead{}
}
\startdata
GMN      & 243 & $1.29\pm0.09$ & $0.22\pm0.01$ & $0.83\pm0.01$ & $12.3\pm1.8$ & $135.78\pm1.63$ & $191.75\pm3.83$ & $4.6\pm0.3$ \\
CAMS     & 10  & $1.35\pm0.17$ & $0.23\pm0.01$ & $0.83\pm0.01$ & $12.8\pm2.0$ & $134.02\pm2.47$ & $194.97\pm3.55$ & $4.4\pm0.4$ \\
EDMOND   & 10  & $1.31\pm0.11$ & $0.22\pm0.01$ & $0.83\pm0.02$ & $12.2\pm0.9$ & $136.12\pm0.66$ & $194.74\pm5.05$ & $4.5\pm0.4$ \\
SonotaCo & 19  & $1.25\pm0.10$ & $0.22\pm0.01$ & $0.83\pm0.01$ & $12.1\pm1.6$ & $135.72\pm1.14$ & \nodata         & $4.6\pm0.3$ \\
Combined & 282 & $1.29\pm0.10$ & $0.22\pm0.01$ & $0.83\pm0.02$ & $12.3\pm1.8$ & $135.78\pm1.66$ & $191.94\pm3.91$ & $4.6\pm0.3$ \\
\enddata
\tablecomments{
All quantities are J2000. Reported values are median values; quoted $\pm$ values are $1\sigma$ dispersions within each subset (not formal uncertainties on the mean). The longitude of ascending node $\Omega$ is not present for SonotaCo.
}
\end{deluxetable*}

\subsection{Search for a tidal disruption signature amongst the meteors}
\subsubsection{Recent tidal signature search (last $<50$~kyr)}
\begin{figure*}[ht!]
\centering
\gridline{
  \fig{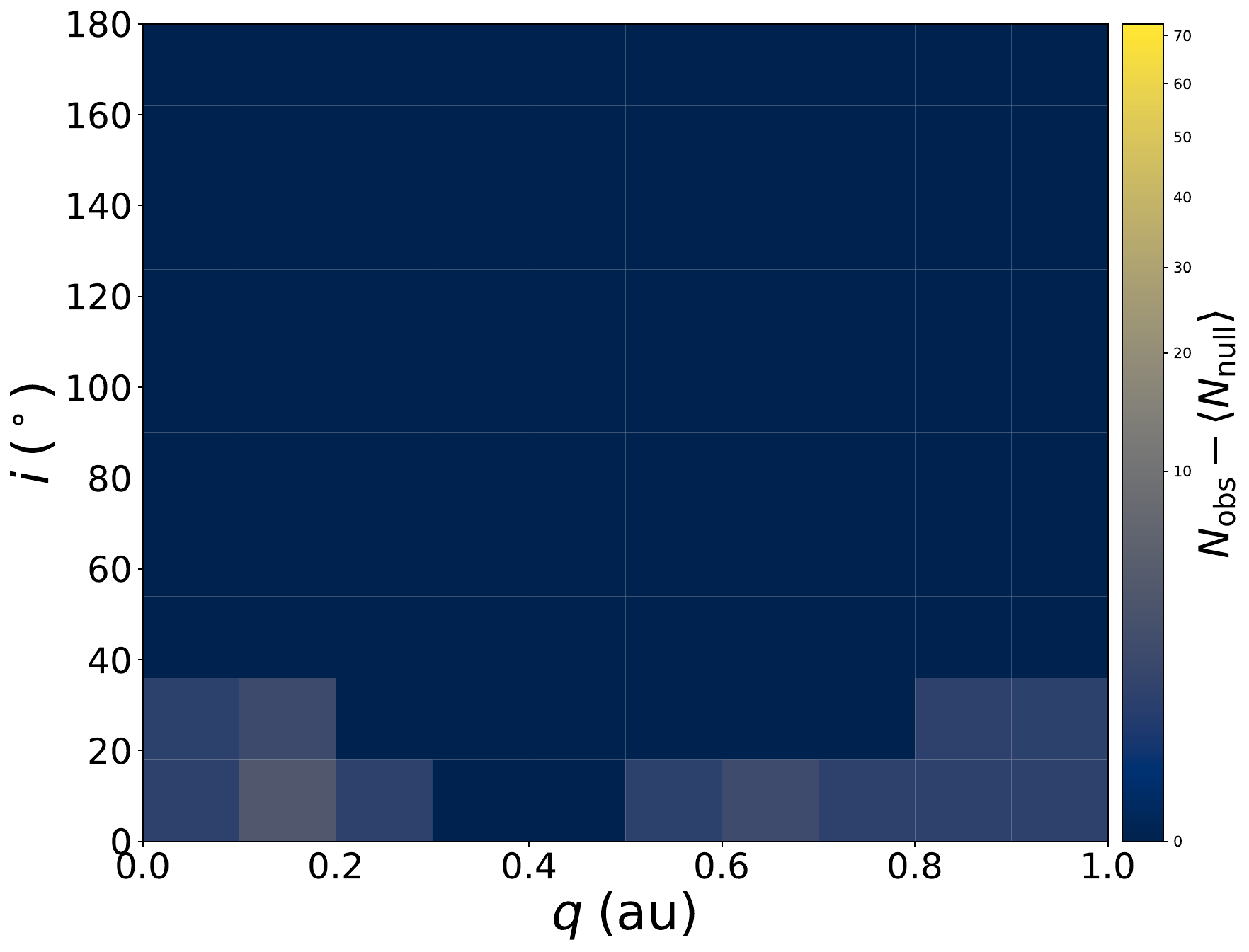}{0.45\textwidth}{(a) Sonotaco}
  \fig{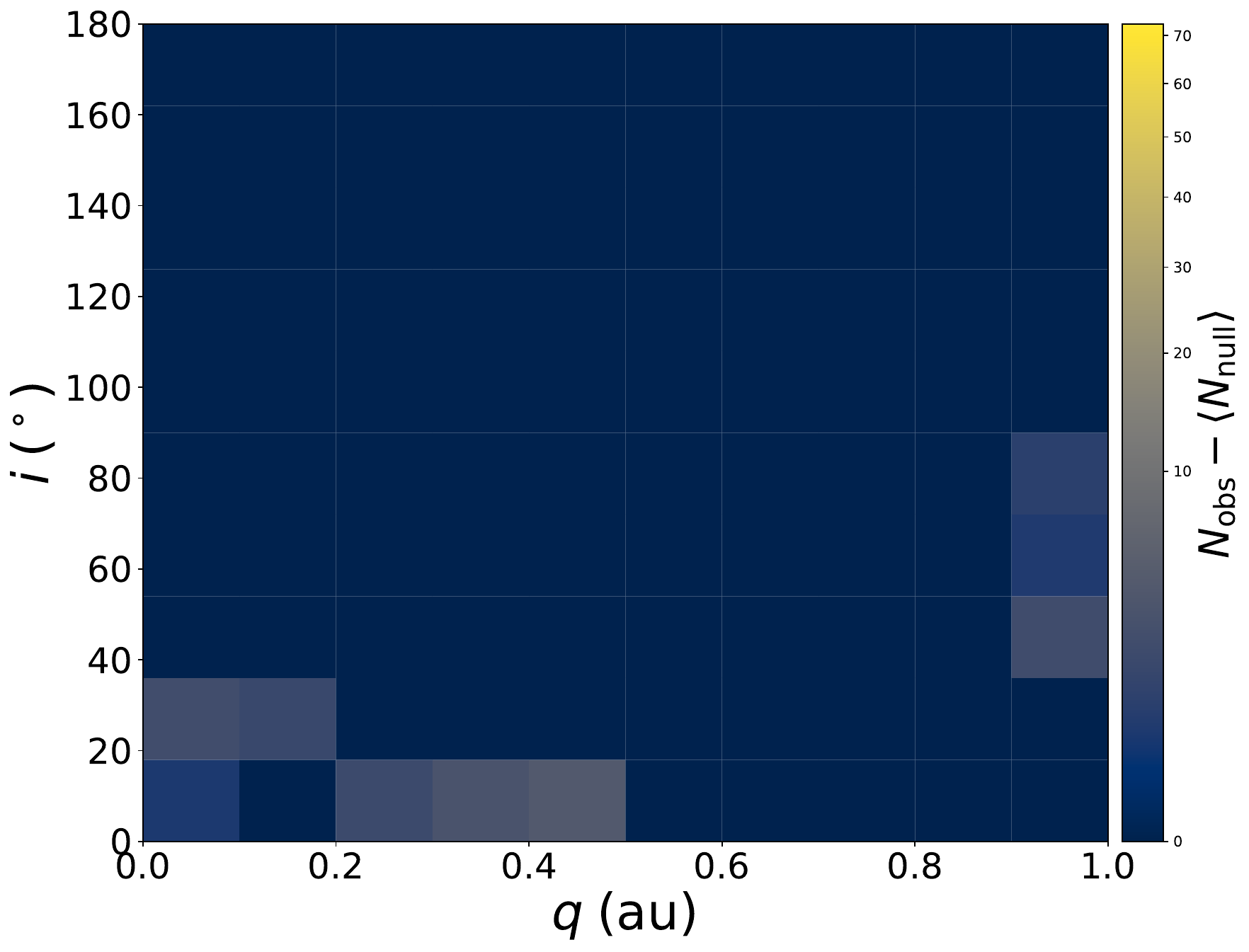}{0.45\textwidth}{(b) EDMOND}
}
\gridline{
  \fig{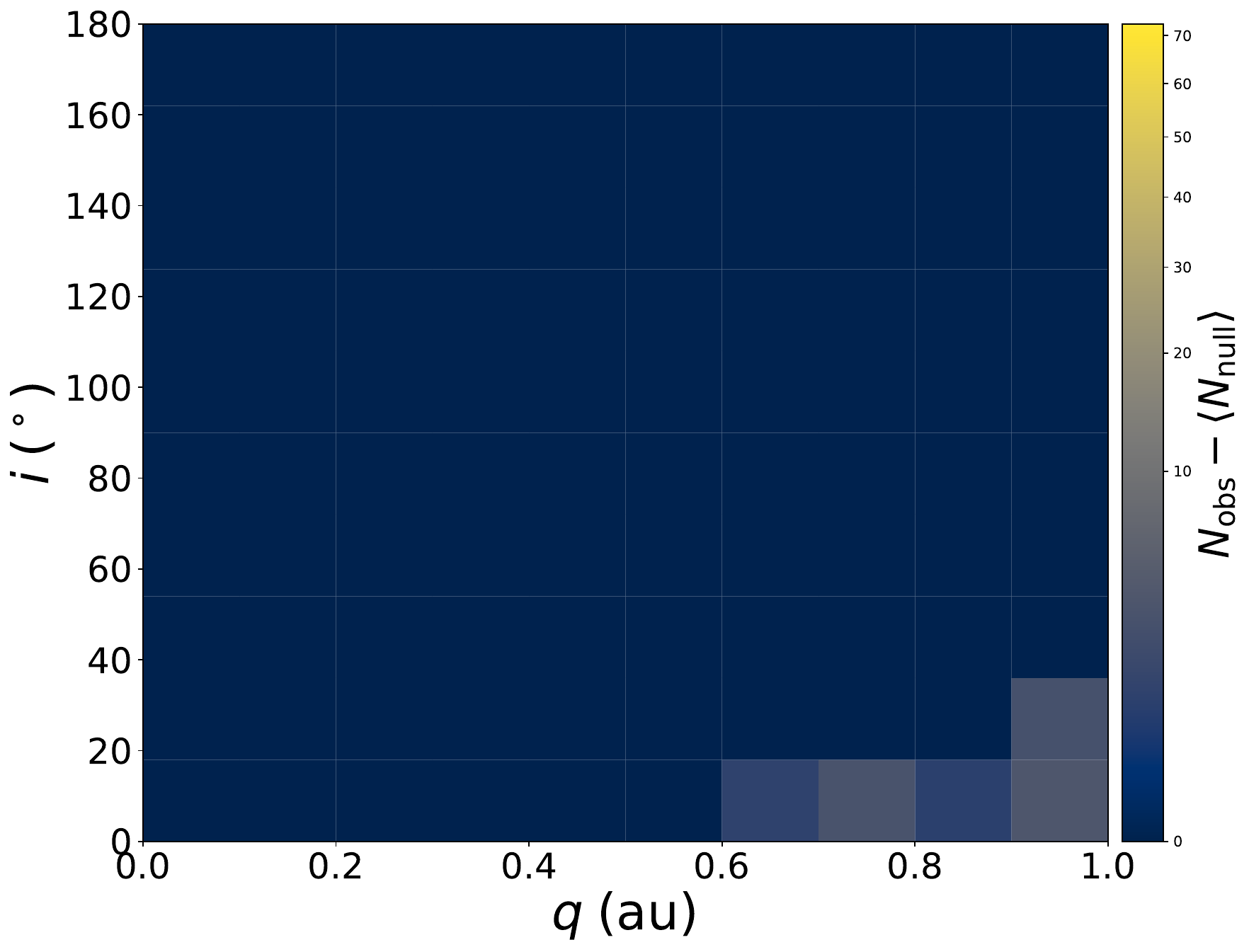}{0.45\textwidth}{(c) CAMS}
  \fig{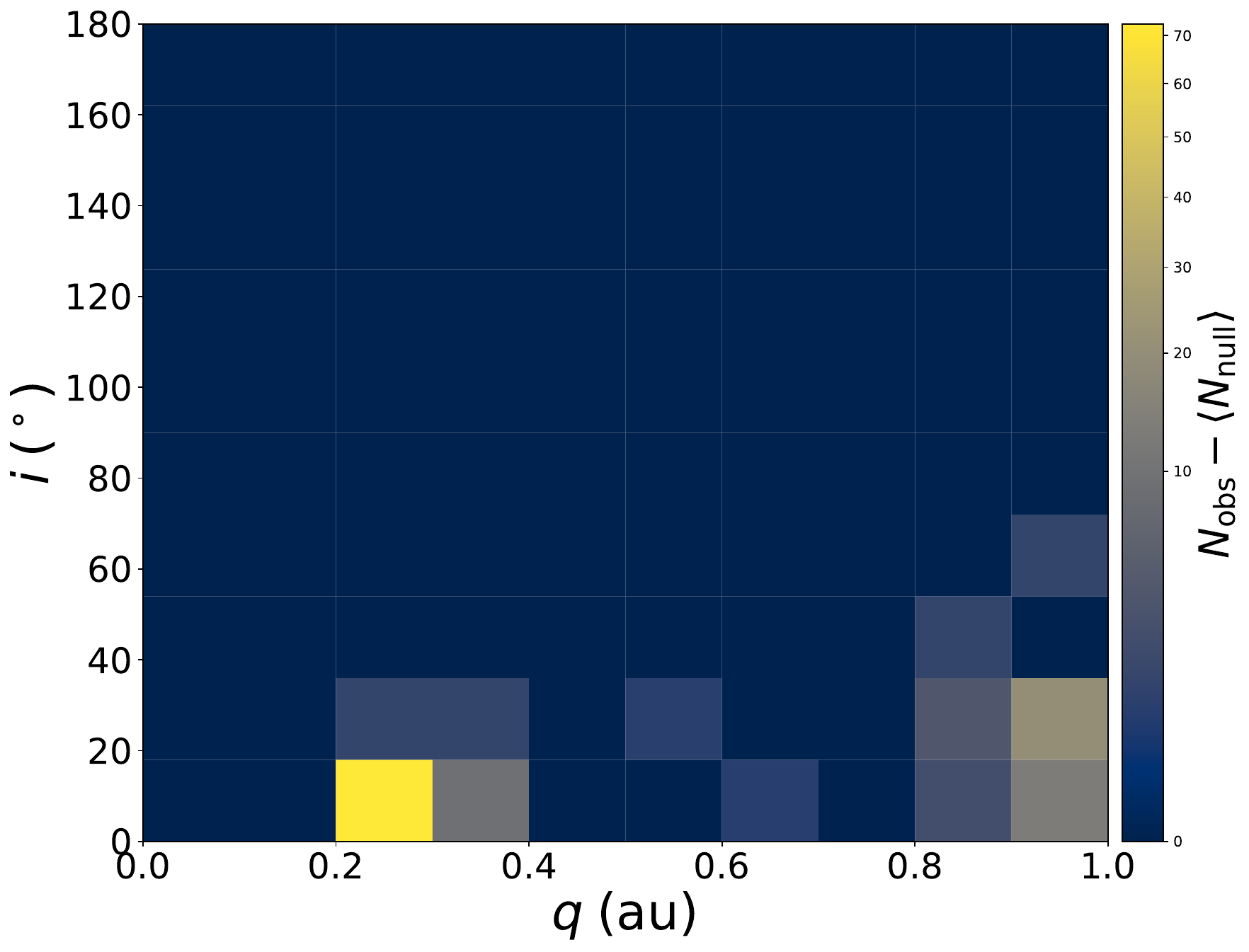}{0.45\textwidth}{(d) GMN}
}
\caption{Perihelion versus inclination (deg) heat map showing bins where there is a statistically significant excess of meteor-meteor pairs where $D_N<0.015$ for (a) Sonotaco, (b) EDMOND, (c) CAMS, and (d) GMN pairs.
\label{fig:stat_sig_2d}}
\end{figure*}

A recent tidal disruption of a near-Earth asteroid is expected to inject a temporally young family of fragments onto closely related orbits, producing an excess of very small-orbit-distance meteor--meteor pairs relative to a randomized null model and, when mapped into orbital-element space, a concentration near Earth-grazing perihelia ($q\approx 1$~au) at low inclinations (e.g., \citealt{granvik2024tidal,schunova2014properties}). To test for such a signature in the video-network meteor populations, we examined the $D_N$-conditioned 2D pair-excess maps (pairs selected with $D_N<0.015$) and their orbit-space projection in $q$--$i$ shown in Figure~\ref{fig:stat_sig_2d}. As described in the figure caption, these panels show where the meteors participating in stable, statistically significant excess bins in the 2D null test (performed in $(U,\lambda_\odot)$) fall in $(q,i)$.

In contrast to the newly identified Virginid shower, any signal in the Earth-like, low-inclination portion of Figure~\ref{fig:stat_sig_2d} is weak and occurs in different $(U,\lambda_\odot)$ bins in different surveys, generally with modest pair surpluses. In GMN, the significant bins that contribute members into the $q\approx 1$~au, low-$i$ region are the $U\in[0.308,0.402]$, $\lambda_\odot\in[324^\circ,342^\circ]$ bin ($N_{\rm obs}=20$ versus $\langle N_{\rm null}\rangle=4.88$; $z=3.35$; surplus $=15.12$ pairs) and the $U\in[0.496,0.590]$, $\lambda_\odot\in[0^\circ,18^\circ]$ bin ($N_{\rm obs}=40$ versus $\langle N_{\rm null}\rangle=13.68$; $z=3.46$; surplus $=26.32$ pairs). In CAMS, the analogous contribution is dominated by the $U\in[0.296,0.392]$, $\lambda_\odot\in[216^\circ,234^\circ]$ bin ($N_{\rm obs}=6$ versus $\langle N_{\rm null}\rangle=0.88$; $z=3.40$; surplus $=5.12$ pairs). In SonotaCo, the only contribution to the $q\approx 1$~au, low-$i$ region comes from a single-pair excess bin at $U\in[0.279,0.374]$ and $\lambda_\odot\in[252^\circ,270^\circ]$ ($N_{\rm obs}=1$ versus $\langle N_{\rm null}\rangle=0.021$; $z=6.92$; surplus $=0.98$ pairs), illustrating that the high-$z$ tail in the smallest networks can be driven by extremely low null expectations rather than by a high absolute pair count. EDMOND shows no meteors from the stable significant-bin membership that fall within $q\in[0.9,1.05]$~au and $i<20^\circ$.

In the absence of a clear ``smoking-gun'' detection of a significant cluster related to tidal disruption, we use the significant-bin member lists associated with Figure~\ref{fig:stat_sig_2d} to place a conservative upper limit on the contribution of any recent tidal disruption to the meteoroid flux represented by these surveys. Defining a tidal-disruption-like candidate region as $q\in[0.9,1.05]$~au and $i<20^\circ$, we count the number of unique meteors within this region among the meteors participating in the stable significant $D_N<0.015$ pair-excess bins. This yields at most 43 meteors in GMN, 8 in CAMS, 2 in SonotaCo, and 0 in EDMOND that could plausibly be attributed to a recent tidal disruption without contradicting the null tests. Relative to the corresponding shower-removed asteroidal sample sizes used in this work (GMN: 122{,}943; CAMS: 60{,}733; SonotaCo: 21{,}254; EDMOND: 30{,}341), these translate to network-level upper limits of $\le 3.5\times 10^{-4}$ (GMN), $\le 1.3\times 10^{-4}$ (CAMS), and $\le 9.4\times 10^{-5}$ (SonotaCo), with no detected contribution in EDMOND at this level. Summed over the combined four-network sample (235{,}271 meteors), this corresponds to a maximum plausible fraction of $\le 53/235{,}271 \simeq 2.3\times 10^{-4}$ of meteors potentially associated with a detectable recent tidal-disruption-like excess in our $D_N<0.015$ significance maps. 

As an independent cross-check for coherent families in the same similarity metric, we compared these results to the DBSCAN clustering in $D_N$ space with $\epsilon=0.03$ and ${\tt min\_samples}=2$ \citep{ester1996density,sugar2017meteor}. The all-network DBSCAN output contains 75{,}291 clustered meteors distributed across 30{,}435 clusters, with the size distribution dominated by pairs (24{,}697 clusters; 81\%), consistent with the expectation that many small, chance-connected components occur at ${\tt min\_samples}=2$ in dense regions of geocentric-parameter space. The most prominent coherent cluster relevant to Figure~\ref{fig:stat_sig_2d} is the previously discussed stream-like cluster, coincident with the dominant GMN low-$q$, low-$i$ excess. In contrast, the candidate tidal-disruption-like meteors identified above do not form a comparably coherent DBSCAN family: the 43 GMN candidates are spread across 22 distinct DBSCAN clusters (largest $N=17$, typically $N=2$), the 2 SonotaCo candidates comprise a single $N=2$ cluster, and the 8 CAMS candidates are either unclustered under DBSCAN or split into only two $N=2$ clusters. Thus, no DBSCAN cluster provides evidence for a survey-robust, stream-like family coincident with the weak $q\approx 1$~au, low-$i$ excess implied by Figure~\ref{fig:stat_sig_2d}.

Taken together, the 2D null-hypothesis maps and the DBSCAN cross-check support the conclusion that the video-network meteor populations do not exhibit a strong, coherent signature of a recent tidal disruption, and that any such contribution is constrained to be a very small fraction of the shower-removed asteroidal sample ($\lesssim 10^{-4}$--$10^{-3}$ per survey, and $\lesssim 2.3\times 10^{-4}$ in the combined sample). This is consistent with rapid decoherence of putative fragment families in meteor-orbit space and/or with the dominance of other disruption pathways in the meteoroid-sized regime \citep{shober2025decoherence}, motivating the complementary population-level comparison of perihelion distributions in the following section.

\begin{figure}[ht!]
\plotone{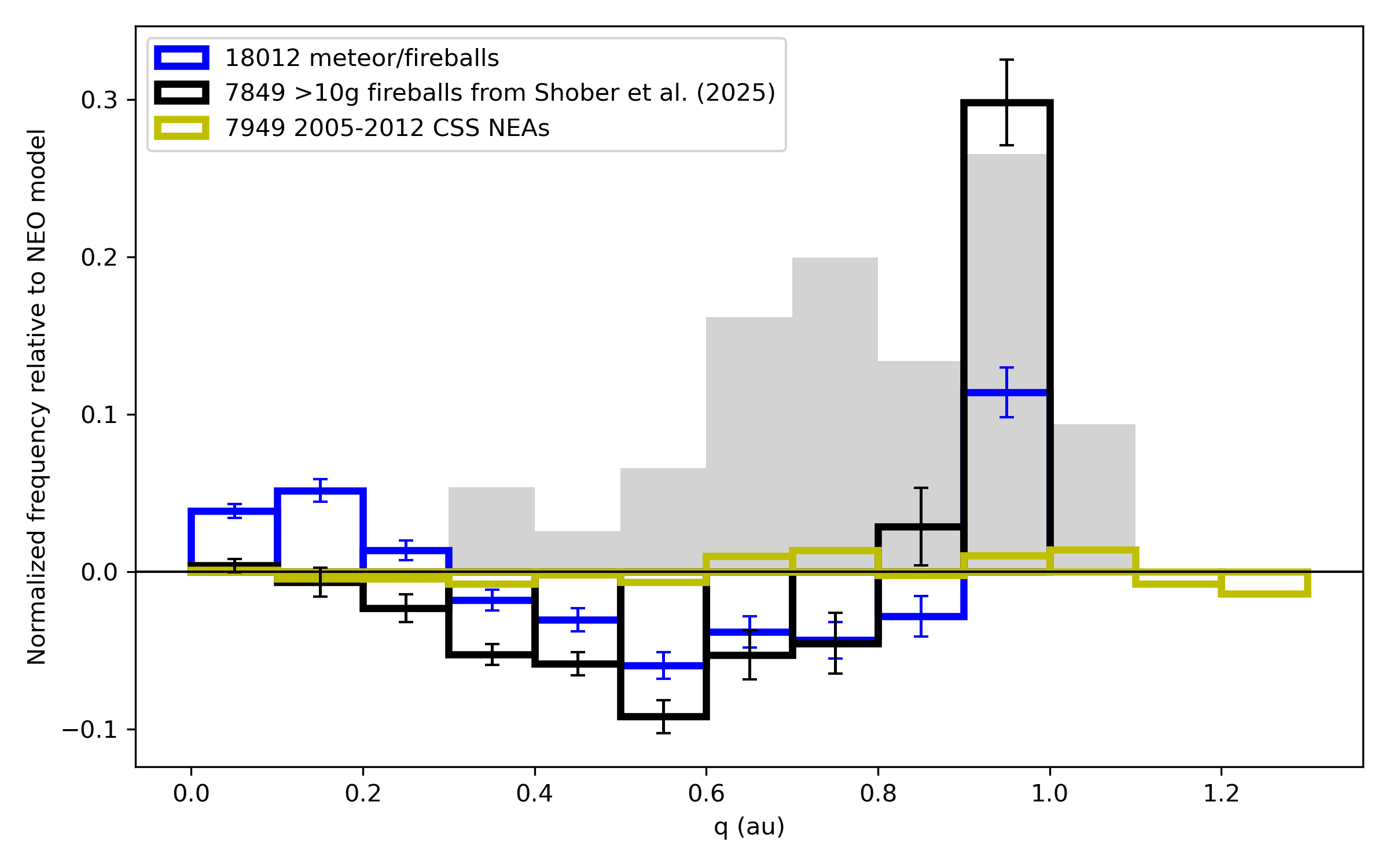}
\caption{Residual perihelion distributions, expressed as the difference between the observed $q$ histogram (weighted by the terrestrial impact probability; see method in \citealp{pokorny2013opik}) and the debiased NEO model of \citet{granvik2016super}. Blue steps show the observation-debiased multi-source perihelion sample used in this work (N=18{,}012; see section~\ref{sec:data}); black steps correspond to the 7\,849 $>10$\,g impacts used to define the top-of-atmosphere population in \citet{shober2025perihelion}; and olive steps represent the 7\,949 near-Earth asteroids discovered by the Catalina Sky Survey during 2005–2012 and analyzed by \citet{granvik2024tidal}. The filled gray histogram reproduces the synthetic $q$ distribution of B-type tidal-disruption fragments from Fig.~1 of \citet{granvik2024tidal}. All histograms are area–normalized to unity. Vertical bars indicate $1\sigma$ uncertainties obtained from bootstrap resampling of each dataset.
\label{fig:g16_comparison}}
\end{figure}

\subsubsection{Long-term tidal signature search (last few million years)}
Similar to the analysis of \citet{granvik2024tidal} and \citet{chow2025decameter}, we also tried to identify unexplained concentrations in the $q$-distribution (Fig.~\ref{fig:g16_comparison}) near the orbits of terrestrial planets that could be reconciled through tidal disruption over long time periods (i.e., millions of years). \citet{chow2025decameter} found no evidence for significant long-term tidal-disruption effects on the meteor distribution based on observations from the EN, USG-sensors, and the Canadian Automated Meteor Observatory. Here, after normalizing the $q$-distribution to that of the model by \citet{granvik2016super}, we find a similar result with a steep rise towards $q\sim1$\,au, with no peak near Venus or Mercury. Additionally, we see a peak of smaller meteoroids ($<1$\,g) on orbits with $q<0.2$\,au. This distribution, nor the distribution from the data used in \citet{shober2025perihelion}, is consistent with that which is expected from a B-type tidal-disruption (50\%–90\% of the total mass is removed; \citealt{richardson1998tidal}), in light gray in Fig.~\ref{fig:g16_comparison}. The long-term signature observed by \citet{granvik2024tidal} is absent. Some process is erasing or overwhelming the signal. As evidenced in Fig.~\ref{fig:g16_comparison}, the meteoroid population, ranging from $\sim10^{-4}$ to $10^{4}$\,kg, appears to have a significantly stronger increase in numbers as it approaches 1\,au and beyond. Despite removing the impact-frequency bias, we still observe a peak near 1\,au because we do not observe meteoroids that, at the very least, do not cross Earth's orbit. 

\begin{figure}[ht!]
\plotone{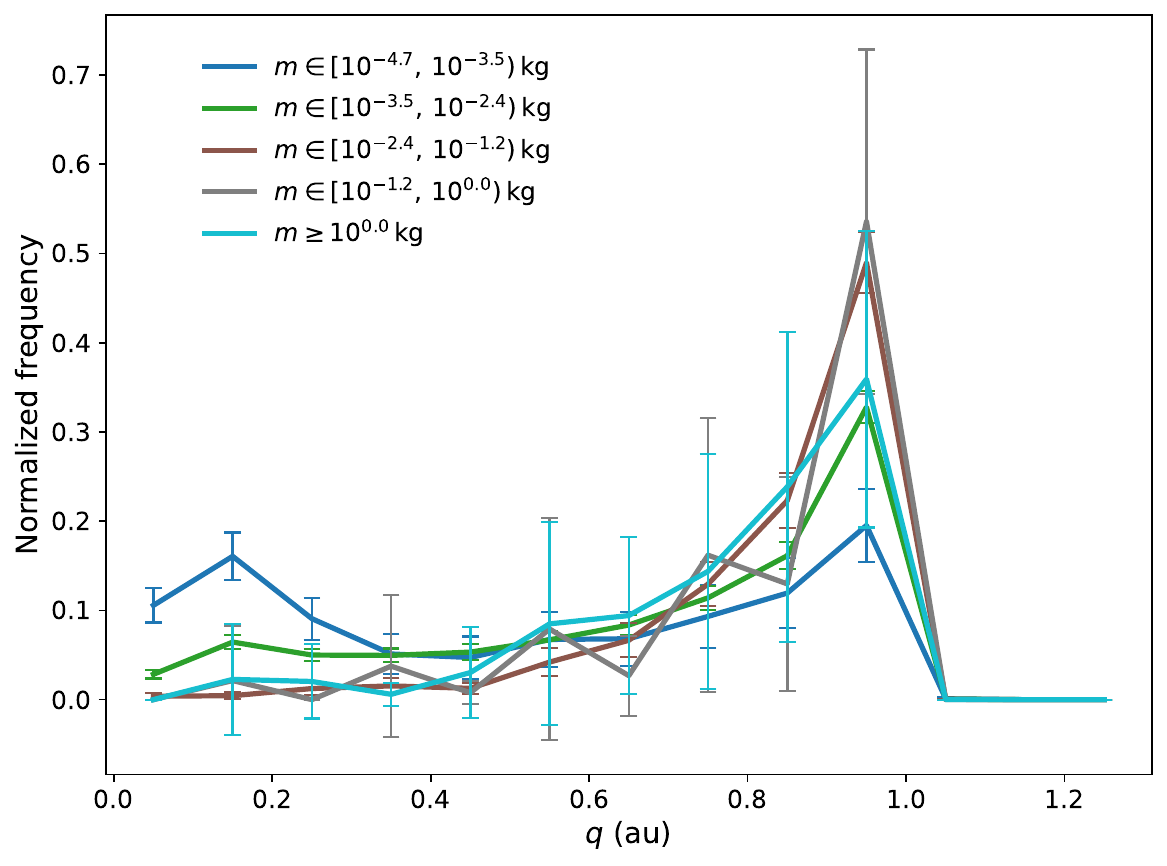}
\caption{Perihelion distribution of the observation-debiased multi-source perihelion sample (N=18{,}012) compiled from CAMS, GMN, EDMOND, GFO, FRIPON, EN, USG sensors, and meteorite falls, weighted by their estimated terrestrial impact probability \citep{pokorny2013opik}. The data are split into five histograms based on the meteoroid's initial mass estimate.
\label{fig:q_distro_normalized}}
\end{figure}

The increased number density relative to the debiased NEA $q$-distribution is attributed to thermal-fracturing-related $q$-filtering, as identified by \citet{shober2025decoherence}. Here ``q-filtering'' denotes a survivorship bias: meteoroids with smaller perihelia are preferentially destroyed/fragmented by temperature-driven processes \citep{capek2010thermal,capek2022ejection}, leaving a relative enhancement at larger $q$. Thermal-stress-driven failure is also expected to depend on meteoroid properties (size, rotation state, and spin-axis geometry) in addition to perihelion distance, implying heterogeneous survival even at fixed $q$. Thermally-induced cracking of boulders, resulting in exfoliation or preferential crack orientation, has also been observed on asteroids Bennu, Ryugu, and the moonlet Dimorphos \citep{molaro2020situ,molaro2020thermal,cambioni2021fine,delbo2022alignment,lucchetti2024fast,schirner2024aligned}. This same process explains the strong correlation between $q$ and the ability of meteoroids to survive passage through the terrestrial atmosphere as meteorites \citep{shober2025perihelion}. Meteoroids are more likely to have been thermally-fragmented into a smaller size-range as $q$ decreases, explaining the increased concentration towards larger $q$ values within the meteoroid population because they are less likely to have been thermally fractured, as seen in Fig.~\ref{fig:q_distro_normalized}. 

To test this theory, we also examined how the $q$-distribution varies as a function of mass range as seen in Fig.~\ref{fig:q_distro_normalized}. As mass decreases, the excess of meteoroids relative to the debiased NEA population at larger $q$ values declines. This trend is consistent with cyclic thermal stresses, governed chiefly by contrasts in thermal-expansion coefficient and Young’s modulus between adjacent mineral grains \citep{molaro2015grain}, efficiently fracture centimeter‑scale bodies and feed debris into the gram bin, yet become progressively less effective below that scale. Once we step into the sub‑gram interval (m$<$1\,g) a new maximum emerges at q$<$0.2\,au. We identify this feature with the ``super‑catastrophic'' regime of \citet{granvik2016super}, in which extreme heating close to the Sun or high-velocity micrometeoroid impacts \citep{wiegert2020supercatastrophic} disrupt objects into millimeter‑sized fragments. The same low‑q enhancement is seen in radar data \citep{wiegert2020supercatastrophic}. Recent irradiance experiments demonstrate that CI-like carbonaceous material can be destroyed on very short timescales (within minutes) at irradiances corresponding to heliocentric distances of order 0.2~au. ``Thermal erosion'', originating from the decomposition of organics and minerals, as the dominant process, is sufficiently effective to account for the lack of carbonaceous asteroids and the abundance of sub-gram meteoroids \citep{tsirvouils2026instantaneous}. This thermal destruction scenario is further supported by meteor spectra \citep{borovivcka2005survey,kasuga2006thermal,matlovivc2019spectral,abe2020sodium,matlovivc2022hydrogen} and thermophysical modeling of NEAs \citep{marchi2009heating,masiero2021volatility}, which reveal a significant depletion of volatile species (e.g., Na and H) in meteoroids with $q < 0.2$~au, suggesting that intense solar heating is actively altering their composition, depletion of volatile species (loss of Na/H-bearing phases) and thermal alteration, and potentially releasing dust in the process.

Independent evidence also comes from the zodiacal‑dust study of \citet{jenniskens2024lifetime}, who measured orbital dispersions and the magnitude size-distribution index $\chi$ for 487 streams. They emphasize two distinct perihelion-dependent effects: (i) dynamical stream dispersion grows faster at smaller perihelion distance, approximately $\propto 1/q$, implying shorter coherence times for low-$q$ streams; and (ii) when restricting the sample to narrow age groups (age inferred from dispersion), $\chi$ increases toward smaller $q$, with an empirical $\chi \propto 1/\sqrt{q}$ dependence. They interpret this shallow perihelion dependence as evidence that, for $0.3<q<1.02$~au, the physical survival of mm--cm meteoroids is limited primarily by temperature-linked processes (thermal stresses), while for $q\lesssim 0.3$~au sublimation of organic compounds can become lifetime-limiting; at still smaller $q$ they note loss of Na-bearing phases ($q<0.08$~au) and, at extreme perihelia, sublimation of refractory core minerals ($q<0.04$~au). In comparison, classical collisional-equilibrium models predict a steeper $1/q$ relationship for centimeter-scale grains \citep{grun1985collisional}. In other words, collisional grinding sets the ceiling on how long cm‑scale meteoroids can survive near Earth, but the progressive shortening of lifetimes toward smaller $q$ must be carved out by temperature‑dependent fragmentation and organic/mineral sublimation.

Thus, the mass‑dependent fade of the large‑q excess and the appearance of a distinct low‑$q$ peak in our data are naturally explained by a hand‑off from thermal fragmentation (dominant for 0.3~$<q<$~1.0\,au) to ``super‑catastrophic'' break‑up driven by thermally-driven erosion ($q<$0.2\,au) \citep{granvik2016super,wiegert2020supercatastrophic,jenniskens2024lifetime,shober2025perihelion,tsirvouils2026instantaneous}. Secondary agents such as hyper‑velocity impacts or YORP‑induced spin‑up may modulate the distribution, but they are not likely required to reproduce the primary ``long-term'' structure seen in Fig.~\ref{fig:q_distro_normalized}.

\section{Conclusions}\label{sec:conclusions}

We characterized the statistical significance of orbital similarity using survey-matched null-hypothesis testing and modern multi-network video-meteor catalogs, with a parallel comparison to NEA-orbit statistics, to search for (i) localized debris structures consistent with recent asteroidal activity and (ii) signatures consistent with recent or long-term encounter-driven tidal disruption. Using cumulative similarity distributions and $D_N$-conditioned ($D_N<0.015$) 2D pair-excess maps in the $(U,\lambda_\odot)$ plane, we detect a new low-perihelion stream and place quantitative upper limits on the fraction of meteors plausibly attributable to a detectable recent tidal-disruption contribution. Here is a summary of our results:
\begin{itemize}
\item After removing shower members and cometary interlopers and generating survey-specific KDE null realizations, we analyzed 235{,}271 meteors/fireballs from GMN (122{,}943), CAMS (60{,}733), EDMOND (30{,}341), and SonotaCo (21{,}254).

\item The dominant localized excess occurs in GMN with $z=6.32$ ($N_{\rm obs}=135$ pairs versus $\langle N_{\rm null}\rangle=38.94$), and maps to a low-$q$, low-$i$ structure. This corresponds to a Dunn--\v{S}id\'ak family-wise $p_{\rm global}=5.2\times10^{-8}$, corresponding to a 5.3~$\sigma$ detection. DBSCAN clustering ($\epsilon=0.03$, minimum samples $=2$) isolates a connected cluster of $N=282$ meteors dominated by GMN (243) but with additional SonotaCo (19), CAMS (10), and EDMOND (10) members. The combined median orbit is low-perihelion ($q=0.22\pm0.01$ au, $i=12.3^{\circ}\pm1.8^{\circ}$, $T_J=4.6\pm0.3$; Table~\ref{tab:mdc_mean_svir_orbit}), with a wide activity window of $20.5^{\circ}$ in solar longitude (Table~\ref{tab:mdc_mean_svir_activity}). The stream's low-$q$ orbit and asteroidal $T_J$ are consistent with activity mechanisms expected for near-Sun asteroids, including thermal fatigue/fracture, dehydration or mineral decomposition, and other thermo-mechanical processing. We have not yet identified a specific parent body, and future observations with telescopes, like NASA's \emph{NEO Surveyor} mission, will be critical for future characterization. The stream has been added by the IAU Meteor Data Center under the provisional name M2026-A1 on the working list.

\item In the Earth-like, low-inclination region often discussed as a potential recent tidal-disruption signature ($q\sim1$ au and $i\lesssim20^{\circ}$), the 2D maps show only weak, survey-dependent excesses, and DBSCAN does not reveal a coherent, survey-consistent family.

\item Counting unique meteors that participate in stable significant 2D bins within $q\in[0.9,1.05]$ au and $i<20^{\circ}$ yields at most 53 meteors across the four-network sample, implying a combined upper limit of $\leq53/235{,}271$ ($\leq2.3\times10^{-4}$) for meteors plausibly attributable to a \emph{detectable} recent tidal-disruption contribution under the adopted thresholds.

\item Impact-probability-weighted perihelion distributions for the observation-debiased multi-source sample (N=18{,}012; CAMS/GMN/EDMOND/EN/FRIPON/GFO/USG sensors and meteorite falls) rise steeply toward $q\approx1$ au relative to a debiased NEO model, lack the Venus-centered peak expected from long-term tidal disruption, and show a sub-gram excess at $q<0.2$ au, consistent with low-perihelion near-Sun catastrophic disruption driven through thermally-driven erosion \citep{tsirvouils2026instantaneous}.

\item The NEA catalog examined here (35{,}012) exhibits a statistically significant excess of very low-$D_H$ pairs. Table~\ref{tab:nea_pairs} lists the ten tightest pairs with $D_H<0.005$; based on the adopted null estimate, roughly half are expected to be false positives, and multiple formation mechanisms (including tidal encounters, YORP-driven rotational fission, binary dissociation, and impacts) must be considered.
\end{itemize}

Future work will focus on dynamical modeling of the new stream (formation age, dispersion, and delivery efficiency) and on controlled parent-body association searches that incorporate orbit uncertainties and the same null-testing framework used here. Additionally, physical constraints on stream and parent candidates, via meteor ablation/fragmentation modeling and, where available, spectra, together with targeted follow-up observations of nearby low-$q$ asteroids.

\bibliography{sample631}{}
\bibliographystyle{aasjournal}

\section{Acknowledgments}
Research was sponsored by the National Aeronautics and Space Administration (NASA) through a contract with ORAU. The views and conclusions contained in this document are those of the authors and should not be interpreted as representing the official policies, either expressed or implied, of the National Aeronautics and Space Administration (NASA) or the U.S. Government. The U.S.Government is authorized to reproduce and distribute reprints for Government purposes notwithstanding any copyright notation herein.

The author would also like to deeply thank Peter Brown, Ian Chow, and Denis Vida for their meaningful discussions during the manuscript's writing.

\end{document}